\documentclass[12pt]{article}

\usepackage{amsmath}
\usepackage{amsfonts}
\usepackage{amssymb}
\usepackage{amsthm}
\usepackage{enumitem}
\usepackage{fancyhdr}
\usepackage{subfig}
\usepackage{float}
\usepackage{bbm}
\usepackage{ulem}
\usepackage{setspace}
\usepackage[export]{adjustbox}
\usepackage{natbib}
\usepackage[linesnumbered,ruled]{algorithm2e}

\usepackage{xr}
\makeatletter
\newcommand*{\addFileDependency}[1]{
  \typeout{(#1)}
  \@addtofilelist{#1}
  \IfFileExists{#1}{}{\typeout{No file #1.}}
}
\makeatother

\newcommand*{\myexternaldocument}[1]{%
    \externaldocument{#1}%
    \addFileDependency{#1.tex}%
    \addFileDependency{#1.aux}%
}
\myexternaldocument{supplement}

\usepackage{graphicx}
\usepackage[colorlinks=true,linkcolor=blue,citecolor=blue]{hyperref}%

\newcommand{\blind}{1}

\makeatletter
\g@addto@macro{\UrlBreaks}{\UrlOrds}
\makeatother

\newtheorem{definition}{Definition}[section]
\newtheorem{proposition}{Proposition}[section]

\newcommand{\R}{\mathbbm{R}}
\newcommand{\g}{\gamma}

\usepackage[usenames,dvipsnames]{color}

\begin{document}

\def\spacingset#1{\renewcommand{\baselinestretch}%
{#1}\small\normalsize} \spacingset{1}




\if1\blind
{
  \baselineskip=28pt \vskip 5mm
    \begin{center} {\LARGE{\bf Elastic depths for detecting shape anomalies in functional data}}
    \end{center}

    \baselineskip=14pt \vskip 10mm

    \begin{center}\large
    Trevor Harris\footnote{\baselineskip=12pt Department of Statistics, University of Illinois at Urbana-Champaign},
        J. Derek Tucker\footnote{\baselineskip=12pt Sandia National Laboratories, Albuquerque, NM},
        Bo Li$^1$,
        Lyndsay Shand$^2$
    \end{center}
    \baselineskip=19pt \vskip 15mm \centerline{\today} \vskip 6mm
} \fi

\if0\blind
{
  \bigskip
  \bigskip
  \bigskip
  \begin{center}
    {\LARGE\bf Elastic depths for detecting shape anomalies in functional data}
\end{center}
  \medskip
} \fi

\medskip
\begin{abstract}
We propose a new family of depth measures called the elastic depths that can be used to greatly improve shape anomaly detection in functional data. Shape anomalies are functions that have considerably different geometric forms or features from the rest of the data. Identifying them is generally more difficult than identifying magnitude anomalies because shape anomalies are often not distinguishable from the bulk of the data with visualization methods. The proposed elastic depths use the recently developed elastic distances to directly measure the centrality of functions in the amplitude and phase spaces. Measuring shape outlyingness in these spaces provides a rigorous quantification of shape, which gives the elastic depths a strong theoretical and practical advantage over other methods in detecting shape anomalies. A simple boxplot and thresholding method is introduced to identify shape anomalies using the elastic depths. We assess the elastic depth's detection skill on simulated shape outlier scenarios and compare them against popular shape anomaly detectors. Finally, we use hurricane trajectories to demonstrate the elastic depth methodology on manifold valued functional data. 
Supplementary materials, including additional simulations, data examples, and an R-package are available online.
\end{abstract}

\noindent%
\vfill

\bigskip
\doublespacing
\section{Introduction} \label{introduction}

\begin{figure}[H]
    \begin{center}
    \includegraphics[width=0.60\textwidth,valign=c]{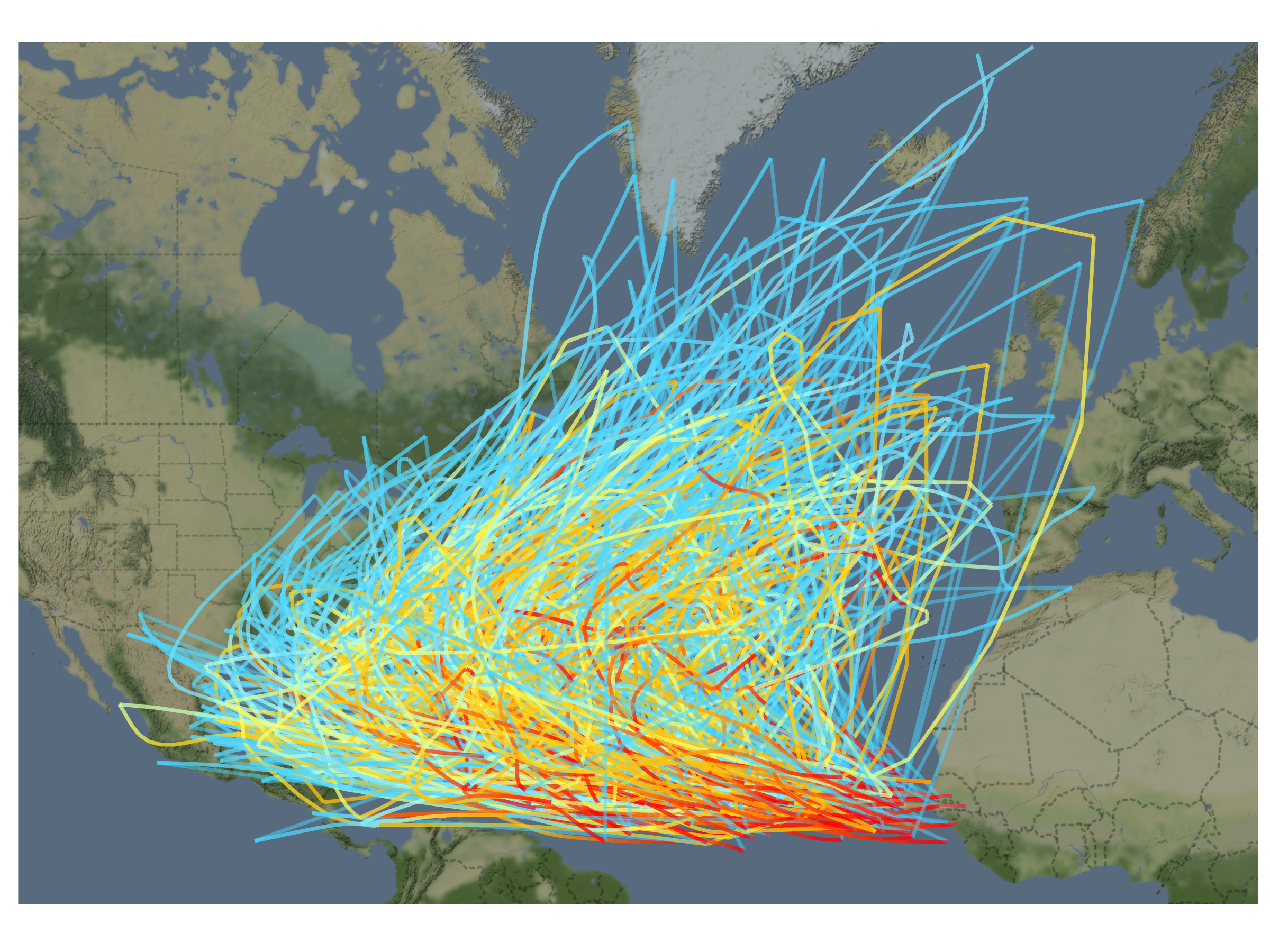}
     \caption{Hurricane trajectories from the HURDAT2 database. Trajectories are colored from red to blue, with red indicating the origin of the hurricane. The familiar ``U-shaped'' pattern emerges as hurricanes generally start near the coast of Africa, migrate to North America, and then curve back towards Europe. This type of data poses many challenges to existing shape outlier methodology. Trajectories are multivariate (latitude and longitude), exist on the surface of a nonlinear manifold ($\mathbbm{S}^2$), and exhibit significant phase and magnitude variability independent of their shape.}
    \label{fig:hurricanes}
    \end{center}
\end{figure}

As data collection methods rapidly advance, functional data and functional data analysis (FDA) have become more prevalent. Functional data refers to data collected continuously across a compact domain, such as a fixed length of time or region of space, and where an observation is an entire curve or surface over the domain, rather than a single value. Examples of functional data include growth rate curves, electrocardiogram (ECG) data, temperature profiles, imaging data containing geometric shapes, and hurricane trajectories (See Figure \ref{fig:hurricanes}).

As with traditional data analysis methods, it is critical to perform exploratory data analysis with functional data. Exploratory analysis can reveal significant trends or anomalies, which could bias post-processing analysis such as model fitting.
Functional anomalies are of particular interest because of the adverse effects they can have on statistical models. Functional anomalies can also be interesting in their own right and can even be the primary focus of study.

The hurricane trajectories in Figure \ref{fig:hurricanes} present a challenge to current functional shape anomaly detectors. Due to the vast distances hurricanes travel, their trajectories resemble paths along the surface of a sphere. Existing methods do not handle spherical valued data, or more generally manifold valued data, so they have to approximate these paths with two-dimensional trajectories. This approximation distorts the distances between curves and consequently has a strong influence on the detection of anomalies. Furthermore, exploratory analysis shows that these trajectories exhibit substantial phase variability that could affect shape outlier detection \citep{srivastava2011}. 

In the functional data setting, identifying functional anomalies, i.e. identifying an entire function as an outlier, is not as straightforward as identifying univariate outliers with visualization methods.
By definition, a functional anomaly is a function that is significantly more ``extreme'' in its characteristics than the rest of the functional data.
Generally, functional anomalies are categorized into two types: magnitude and shape anomalies \citep{dai2018multivariate}. 
Magnitude anomalies are functions that clearly lie outside of the range of all other functions and are usually detected through data visualization methods \citep[e.g.,][]{hyndman2010, sun2011, myllymaki2017}. On the other hand, shape anomalies take on a different shape or pattern than the rest of the data. They are more challenging to identify with visualization methods because they can lie hidden amongst the rest of the functions \citep{arribasgil2014}.
Examples of shape anomalies include trajectories with more or less curvature, trajectories with more or fewer oscillations, or trajectories sampled from a process having a different mean function than the rest of the data.
These are only a few examples; the possible ways a trajectory can be shape outlying are innumerable. Furthermore, the information contained in the shapes of curves ``matters a great deal'' \citep{horvath2012inference} and can be quite different from the information contained in their magnitudes. Therefore, it is important to develop methods that can isolate shape information and identify shape outliers from the rest of the data. 

Many methods for identifying shape outliers rely on the notion of functional data depth. Functional data depth is a family of methods used to define centrality and induce a center-outward ordering on the sample functions \citep{liu1999multivariate, zuo2000, mosler2012general}. All depth measures rank functions from most central (higher depth values) to least central (lower depth values) and are typically scaled to take on values in $[0, 1]$.

Many outlier detection methods based on depth decompose total depth (or outlyingness) into magnitude and shape depth (or outlyingness). \citet{arribasgil2014} proposed the Outliergram, a visualization tool for the shape and magnitude components of trajectories, using the half-space depth \citep{tukey1977} and the band depth \citep{lopez2009}. Later, \citet{huang2019decomposition} introduced the total variation depth, which they decomposed into magnitude and shape components. In \citet{dai2019}, the authors pointed out that these methods, which rely on integrated depth, do not efficiently represent the centrality of functions. To remedy this, \citet{rousseeuw2018} and \citet{dai2019} simultaneously proposed the concept of directional outlyingess. Directional outlyingness has since been used as the basis of the functional outlier map \citep{rousseeuw2018} and the magnitude-shape plot tool \citep{dai2018multivariate}. Both of these methods again decompose their depth measures into magnitude and shape components, for the separate identification of magnitude and shape anomalies.

Other outlier detection methods proposed in the recent literature account for the geometry of the functions. \citet{kuhnt2016} developed the functional tangential angle (FUNTA) pseudo-depth based on the tangential angles of the intersections of the centered data. \citet{nagy2017} proposed two modifications of previous depth notions to better identify shape outliers by emulating derivatives with multidimensional projections. \citet{xie2017geometric} separated the variability of functional data into amplitude and phase components, using the registration methods of \citet{srivastava2011}, and displayed this variability using independent boxplots for each component.  \citet{xie2017geometric} showed how treating the phase and amplitude components of trajectories separately could greatly improve the detection rate of shape outliers. Their method, however, falls short of fully characterizing the shape distribution and instead relies on an optimization procedure to partially approximate the boundary of the inlier distribution.

Finally, there are methods based on Functional Principal Component Analysis (FPCA), which extract the features of normal shapes and detect anomalies by finding functions with abnormal features. These methods include a Step-wise Functional Outliers Detection test \citep{yu2012outlier} and an FPCA score based distance test \citep{ren2017projection}. FPCA based feature extraction can be quite powerful for detecting many types of shape anomalies, but FPCA is also known to be deficient when temporal variability is present \citep{srivastava2011, tucker2013}. Furthermore, these methods are not designed to explicitly identify shape anomalies separately from magnitude anomalies. Because we assume that both temporal variability and magnitude variability are present and are nuisance properties that mask shape, the problems and data considered here are quite different from the ones considered by FPCA methods. For those reasons, these methods were not included in the comparisons in Section \ref{sec:multi}.

In this paper, we introduce a new family of depth measures, called the elastic depths, based on the elastic shape distances used in \citet{xie2017geometric}. We first use the elastic distances to directly measure the distance between the shapes of individual trajectories. This allows us to define a notion of data depth that appropriately captures the distribution of the trajectories' shapes and allows for the explicit identification of shape outliers. Current literature falls short of fully representing the shape distribution and instead uses either a surrogate for shape or only approximates the shape distribution. Explicitly using the shape distribution, as we do in this paper, also confers several practical benefits, namely
\begin{enumerate}
    \item Finding the inner quartile regions and outlier bound is trivial due to our depth based representation. The outlier bound does not require any optimization procedures and has only a single input parameter. We demonstrate in Section \ref{sec:sensitivity} of the online supplement that the performance of our method is relatively insensitive to the value of the input parameter.
    
    \item In Sections \ref{sec:single} and \ref{sec:multi}, we empirically demonstrate the highly competitive and often superior outlier detection skill of the elastic depth based boxplots. We show that the elastic depths are consistently the top performing detection method across many different shape classes and that they can identify shape outliers in the presence of substantial translation and phase noise.
    
    \item Shape distributions can easily be generalized to manifold valued trajectories, such as the hurricane trajectories (Figure \ref{fig:hurricanes}), because the elastic depths are based purely on distance. Different manifolds merely require different distance metrics. In Section \ref{sec:hurricanes}, we show the elastic depths applied to the Hurricane trajectories and in Section \ref{sec:examples_appendix} of the online supplement we show additional examples of the elastic depths applied to $\mathbb{R}$ and $\mathbb{R}^2$ valued data.
\end{enumerate}


\section{Background} \label{background}
\subsection{Data Depth}
Data depth is a general notion of measuring the centrality of observations with respect to a distribution. In the FDA literature, data depth is the dominant method used to define centrality and induce ordering on a function space. Given a distribution $P$ on a function space $F$, a depth function maps each trajectory $f \in F$ to a value in $[0, 1]$, such that the closer a trajectory is to the center of $P$, the higher its depth value is. If this mapping is monotonic, that is, higher depth values necessarily mean higher centrality, then the depth function induces a center-outward ordering on the function space $F$ with respect to $P$. This makes depth a natural framework for evaluating the outlyingess of observations. High depth values mean an observation is very close to the center of $P$ so, conversely, low depth values mean an observation is very far from the center of $P$. Therefore, trajectories with extremely low depth values, as compared to the rest of the distribution, are likely to be outlying or anomalous.


\subsection{Elastic shape analysis}\label{sec:esa}


Elastic shape analysis (ESA) is a collection of techniques for registering functional data through a phase-amplitude separation procedure and for performing statistical analysis on the separated phase and amplitude components \citep{srivastava2011, kurtek2011Signal, tucker2013}.
Phase and amplitude represent two orthogonal components of a function's variability. The amplitude component represents variability in shape, where shape refers to the properties of a function that remain unchanged under the shape preserving transformations: rotation, translation, scaling, and phase \citep{srivastava2016functional}. The phase component represents the ``domain'' or ``timing'' variability of the trajectories. Because amplitude is invariant to these phase transformations, amplitude is distinct from the usual concept of magnitude. Magnitude measures the size of the observed realization of a trajectory while amplitude measures the size of the trajectories shape.

The distinguishing feature of ESA is the use of the Square Root Slope Function (SRSF) for registration \citep{kurtek2011Signal}. For real valued trajectories, the SRSF bijectively maps, up to an additive constant, a real valued function $f$ to its normalized gradient $f' / \sqrt{|f'|}$. Under ESA, two real valued trajectories are registered by elastically deforming the domain of one function such that the $L^2$ distance between the SRSFs of the two functions is minimized (Section \ref{sec:r1}).
The amount of elastic deformation needed to register two functions is measured by the phase distance (Section \ref{sec:elastic_definitions}), while the residual $L^2$ distance between the SRSFs, post registration, defines the amplitude distance between them (Section \ref{sec:r1}). Together they are known as the elastic distances. The key insight of ESA is that by registering SRSFs, instead of trajectories directly, the amplitude distances are proper metrics and they are invariant to the shape preserving transformations. Thus, amplitude distance can be used to define the distance between the shapes of functions.

Later the Square Root Slope Velocity Function (SRVF) \citep{srivastava2011} was introduced to register $\mathbbm{R}^n$ valued trajectories and the Transported Square Root Slope Velocity Function (TSRVF) \citep{su2014statistical} was introduced to register Riemannian manifold valued trajectories, such as trajectories observed on the unit sphere $\mathbbm{S}^2$. These notions allow us to calculate amplitude distances between multivariate functions and manifold valued functions respectively.
We present the details for computing amplitude distances for $\mathbb{R}$ valued trajectories in Section \ref{sec:r1}. The details for $\mathbbm{R}^n$ valued and $\mathbbm{S}^2$ valued trajectories are deferred to the Appendix (Section \ref{sec:elastic_definitions}).

The advantages of the ESA approach to shape analysis have previously been shown in the works of \cite{srivastava2011, kurtek2011Signal, tucker2013, su2014statistical}. ESA rigorously defines the shape space for a given class of trajectories and then defines a way to construct a proper distance metric on that shape space. The ESA based metrics are preserved under the shape preserving transformations: translation, scale, rotation, and reparameterization (phase). This improves theoretically over alternative shape metrics, such as \cite{huang2016total, dai2018multivariate, dai2019}, that do not guarantee invariance or equivariance to shape transformations. The ESA framework is also general enough to apply to data observed in $\mathbbm{R}$, $\mathbbm{R}^n$, $\mathbbm{S}^2$, and any Riemannian manifold $\mathcal{M}$ that has an intrinsic metric. This is important for our motivating example, Atlantic hurricane trajectories (Figure \ref{fig:hurricanes}), which are observed on the surface of a sphere.

\subsection{Amplitude distance for $\R$ valued functions} \label{sec:r1}
Let $F_R = \{f:[0, 1] \mapsto \mathbb{R}, f \text{ differentiable} \}$ be the class of differentiable trajectories on $[0, 1]$ mapping to $\R$. The Square Root Slope Function was introduced in \citet{srivastava2011} as the following transformation on trajectories $f \in F_R$:
\begin{definition} \label{def:srsf}
    Let $f$ be a differentiable trajectory in $F_R$, the Square Root Slope Function (SRSF) of $f$ is
    \[
        q_f(t) = \frac{f'(t)}{\sqrt{|f'(t)|}}.
    \]
\end{definition}
As was shown in their paper, the SRSF is a bijective mapping, up to an additive constant, from the space $F_R$ to the space of square integrable functions $L^2$. This means that for two functions $f, g \in F_R$, the norm on $L^2$
\begin{equation} \label{eqn:R1metric}
    ||q_f - q_g||_2 = \sqrt{\int_0^1 |q_f(t) - q_g(t)|^2 dt},
\end{equation}
where $q_f, q_g$ are $f$ and $g$'s associated SRSFs, is a proper distance between $f$ and $g$ themselves. This norm is particularly important for shape analysis because it is phase invariant \citep{srivastava2016functional}. That is, for any phase function $\g \in \Gamma$
\[
||q_{f \circ \gamma} - q_{g \circ \gamma}||_2 = ||q_f - q_g||_2,
\]
where $f \circ \gamma (t) = f(\gamma(t))$, $\forall t \in [0, 1]$. Technical descriptions of phase functions $\g$ and phase space $\Gamma$ are deferred until Section \ref{sec:elastic_definitions}, but $\g$ functions essentially acts to deform the domain $[0, 1]$.

Phase invariance means that Equation \ref{eqn:R1metric} is measuring some quantity that is independent of the representation, or phase, with which two functions are observed. This is only true if $f$ and $g$ share a common phase representation, so in order to find the amplitude distance between two arbitrary $f, g \in F_R$ we need to first place them in phase with each other. That is, we need to find some $\g^* \in \Gamma$ such that 
\[
\g^* = \arg\inf_{\g \in \Gamma}||q_f - q_{g \circ \gamma}||_2,
\]
so that $||q_f - q_{g \circ \g^*}||_2$
measures the difference in their amplitudes. This can be more directly stated by defining the amplitude distance between $f$ and $g$ as in \citet{srivastava2011}:
\begin{definition}{(Amplitude distance)} \label{eqn:R1dist}
Let $f$ and $g$ be two trajectories in $F_R$, then the amplitude distance between $f$ and $g$ is
\[
d_a(f, g) = \inf_{\g \in \Gamma}||q_f - q_{g \circ \g}||_2,
\]
where $q_f$ and $q_{g \circ \g}$ denote the SRSF's of $f$ and $g \circ \g$ respectively.
\end{definition}

\section{Elastic Depth} \label{elasticdepth}

\subsection{Definition of Elastic Depth} \label{defelasticdepth}

The exact analytic form of the elastic depths will greatly depend on the manifold on which the functional objects live. This is because the elastic depths are inherently going to be a robust summary of the distances between functional objects and the definition of distance between functional objects will inherently depend on the manifold on which they are observed.
For instance, the distance between trajectories in $\mathbbm{R}^n$ is very different from the distance between trajectories on $\mathbbm{S}^2$. For the elastic distances to exist, however, it is only required that the data live on a Riemannian manifold (Section \ref{defelasticdepth}), such as $\mathbbm{R}^n$ or $\mathbbm{S}^2$, because the TSRVF of \citet{su2014statistical} can always be used to construct appropriate phase and amplitude distances. Therefore, we will only assume that our data live on a Riemannian manifold $M$ with an intrinsic metric, so the space of functions we consider is defined as
\begin{equation*}
    F_M = \{f:[0,1] \mapsto M, f \text{ is differentiable and M is a Riemannian manifold} \}.
\end{equation*}

The amplitude distance between two functions $f_1, f_2 \in F_M$ will generically be denoted as $d_a(f_1, f_2)$ and the phase distance between them as $d_p(f_1, f_2)$. The exact form of these distances is left unspecified because the amplitude distance is highly dependent on the manifold $M$. See Section \ref{sec:r1} for the definition of amplitude distance for $\mathbb{R}$ valued trajectories and \ref{sec:elastic_definitions} for the definitions of phase distance and amplitude distance for $\mathbbm{R}^n$ valued and $\mathbbm{S}^2$ valued trajectories. 
We now define the elastic depths for data observed on a manifold $M$ using the associated amplitude and phase distances. Let $P$ denote a distribution supported on the space $F_M$ and suppose we observe a function $f \in F_M$. We first introduce the idea of outlyingness, which describes the degree to which $f$ is an outlier relative to $P$. We further divide this concept into amplitude and phase outlyingness, using the amplitude and phase distances respectively. This is done to separately quantify the shape outlyingness and phase outlyingness of $f$ relative to $P$.
Amplitude and phase outlyingness are respectively denoted as $O_a$ and $O_p$ and are defined as
    \begin{align*}
        O_a(f, P) &= \inf_{t \in \R^+} \left\{P( d_a(f, X) \leq t) \geq \frac{1}{2} \right\}, \hbox{and}\\
        O_p(f, P) &= \inf_{t \in \R^+} \left\{P( d_p(f, X) \leq t) \geq \frac{1}{2} \right\}
    \end{align*}
where $X$ is a random function in $F_M$ and drawn from the distribution $P$. The outlyingness functions $O_a$ and $O_p$ robustly summarize the pairwise distances between $f$ and all other functions $X \in F_M$. These two functions define a measure of outlyingness such that if $O_a(f, P)$ is large then $f$ is generally dissimilar in amplitude from other functions $X \in F_M$, with respect to the distribution $P$. Likewise, if $O_a(f, P)$ is small then $f$ is similar in amplitude to other functions $X \in F_M$.

To convert $O_a$ and $O_p$ into depth functions we invert them with the type B depth construction of \citet{zuo2000}: 
\begin{align}
D_a(f, P) &= (1 + O_a(f, P))^{-1}, \\
D_p(f, P) &= (1 + O_p(f, P))^{-1}.
\end{align}
$D_a(f, P)$ and $D_p(f, P)$ are respectively called the amplitude depth and phase depth of $f$ with respect to $P$. Together we denote them the \textit{elastic depths}. The purpose of inverting the outlyingness functions in this manner is to create bounded measures of centrality, i.e. depths, on the amplitude and phase spaces associated with $F_M$. When depths, such as $D_a$ and $D_p$, satisfy the properties outlined in Section 3.2, they provide a non-parametric and moment free characterization of the distribution $P$. Larger depth values indicate higher centrality and low outlyingness, while lower values indicate higher outlyingness. Thus, $D_a$ and $D_p$ provide a simple and rigorous way to identify outliers based on the underlying distribution $P$.

\subsection{Properties} \label{sec:properties}
Within the depth literature there have been many desirable properties discussed for both multivariate and functional data depths; see \citet{zuo2000} and \citet{mosler2012general} for comprehensive reviews. These properties ensure that a depth function properly measures the notion of depth or centrality. For instance, a depth function needs to be location and scale invariant (or equivariant) and it should decrease monotonically from a natural point of symmetry. Since our depth is purely for functional data we concentrate on the central properties of \citet{mosler2012general}. These properties are established for the amplitude depths because amplitude is the primary concern of shape analysis.

The elastic depths are based on proper distance metrics so they inherit certain properties such as translation invariance and scale equivariance automatically. On some manifolds, such as $\R^2$, scale equivariance can be promoted to scale invariance because the trajectories are constrained to live on an $L^2$ ball. Invariance to simultaneous reparameterization (simultaneous phase invariance) was shown in \citet{srivastava2011} for amplitude distances between $\R$ and $\R^n$ valued trajectories and then later extended to $\mathbb{S}^2$ valued trajectories in \citet{su2014statistical}. Consequently, the amplitude depths are also invariant to simultaneous reparameterization.

Other properties, such as phase invariance, maximality of the center, and convex level sets are essential for shape anomaly detection but are not simple corollaries of the amplitude distance. We outline these properties, as they apply to amplitude depth, below. All proofs are deferred to the online supplement Section \ref{sec:proofs}.

\begin{proposition}[Phase invariance] \label{prop:phase_inv}
Let $\Gamma$ be the space of warping, or phase, functions defined in Section \ref{sec:phase} and let $\g \in \Gamma$. Let $F_M$ be the space $M$-valued differentiable functions as in Section \ref{defelasticdepth}, let $f \in F_M$ and suppose $P$ is a distribution supported on $F_M$. Then
\[
D_a(f \circ \gamma, P) = D_a(f, P),
\]
where $D_a(\cdot, P)$ is the amplitude depth of trajectories on $F_M$ with respect to $P$.
\end{proposition}
This property is unique to the elastic depths and ensures that the amplitude depths are invariant to the phase under which each trajectory is observed. This property, in conjunction with translation and scale invariance (equivariance), means that the amplitude depth is invariant to the shape preserving transformations. We can, therefore, say that amplitude depths are appropriately capturing our definition of shape.

\begin{proposition}[Maximality of the center] \label{prop:maximality}
 Let $F_M$ be the space $M$-valued differentiable functions as in Section \ref{defelasticdepth}, let $f \in F_M$ and suppose $P$ is a distribution supported on $F_M$. A trajectory $s \in F_M$ is the amplitude Fr\'echet median of $P$ if and only if $s = \arg\max_{f \in F_M} D_a(f, P)$, where $D_a(\cdot, P)$ is the amplitude depth of trajectories on $F_M$ with respect to $P$.
\end{proposition}

Maximality of the center guarantees that the maximizer of the amplitude depths, denoted the amplitude depth median, is the actual Fr\'echet median of the distribution. The Fr\'echet median is the trajectory that minimizes the median distance between itself and all other points in the space. This property ensures that the amplitude depths start their ordering from the true amplitude center of the distribution.

\begin{proposition}[Convex level sets] \label{prop:convexity}
Let $F_M$ be as in Section \ref{defelasticdepth}, let $f \in F_M$ and suppose $P$ is a distribution supported on $F_M$. Let $D_{a, \alpha}(P) = \{f \in F_M: D_a(f, P) \geq \alpha \}$ be the upper level sets for the amplitude elastic depth for all $\alpha \in [0, 1]$. Then $D_{a, \alpha}(P)$ is a convex set.
Similarly the upper level sets for the phase elastic depth $D_{p, \alpha}(P) = \{f \in F_M: D_p(f, P) \geq \alpha \}$ are convex for all $\alpha \in [0, 1]$.
\end{proposition}

Convexity of the level sets implies that depths decrease monotonically from the center of the distribution. In conjunction with Maximality of the Center, level set convexity guarantees that the elastic depths are measuring centrality in amplitude space and phase space. This property further distinguishes the elastic depths from previous depth notions because they do not directly characterize centrality in the appropriate shape spaces. We use these convex level sets as the theoretical basis for the construction of the depth boxplots (Section \ref{identifyoutliers}) and for depth thresholding (Section \ref{thresholding}).

\subsection{Estimating Elastic Depths} \label{estimation}

As in Section 3.2, let $F_M$ be the space of differentiable functions on the Riemannian manifold $M$ and let $P$ represent a distribution supported on $F_M$. 
Suppose we observe $f_1,...,f_n \sim P$. The amplitude and phase depths of each $f_i$, $i \in 1,...n$, can be estimated empirically using their respective sample outlyingness functions.
The sample amplitude and phase outlyingness functions are respectively denoted as $O_{a,n}$ and $O_{p,n}$ and are defined as:
\begin{align*}
O_{a,n}(f, P_n) &= \text{median}\{d_a(f, f_1), ..., d_a(f, f_n)\}\\
O_{p,n}(f, P_n) &= \text{median}\{d_p(f, f_1), ..., d_p(f, f_n)\},
\end{align*}
where $P_n$ denotes the empirical distribution of the functions $f_1,...,f_n$. Using the same construction as before, we invert the sample outlyingness functions into sample depths
\begin{align}
    D_{a,n}(f, P_n) &= (1 + O_{a,n}(f, P_n))^{-1} \\
    D_{p,n}(f, P_n) &= (1 + O_{p,n}(f, P_n))^{-1},
\end{align}
for amplitude and phase respectively. The following proposition asserts the uniform consistency of this depth estimator. 

\begin{proposition}[Uniform Consistency] \label{prop:consistency}
Let $F_M$ be as in Section \ref{defelasticdepth}, suppose $P$ is a distribution supported on $F_M$, let $f_1,...,f_n \sim P$, and let $P_n$ represent the empirical distribution of the sample. Then
\begin{align*}
    \lim_{n \rightarrow \infty} \sup_{f \in F_M} |D_{a,n}(f, P_n) - D_a(f, P)| &= 0 \\
    \lim_{n \rightarrow \infty} \sup_{f \in F_M} |D_{p,n}(f, P_n) - D_p(f, P)| &= 0,
\end{align*}
where $D_a(\cdot, P)$ is the amplitude depth of trajectories on $F_M$ with respect to $P$, $D_{a,n}(f, P_n)$ is the amplitude depth's empirical counterpart, $D_p(\cdot, P)$ is the phase depth of trajectories on $F_M$ with respect to $P$, and $D_{p,n}(f, P_n)$ is the phase depth's empirical counterpart.
\end{proposition}

\section{Identifying outliers} \label{identifyoutliers}
Data depth is a natural framework for outlier detection because it provides a center-outwards ordering of the data. Functions with very low depth values are strong candidates for outliers because they are statistically far from the center of the distribution. As mentioned in Section \ref{introduction}, there have been many methods, many based on functional depth in some way, for detecting shape anomalies proposed in the literature. These methods typically construct an outlier cutoff boundary on either the depths or the functions and classify any trajectory as an outlier if it exceeds these bounds. In the next two sections, we introduce two simple ways of defining an outlier cutoff point based on elastic depth.

\subsection{Depth Boxplots} \label{boxplots}
The first method we introduce is called the Depth Boxplot, which is a half-boxplot constructed on the elastic depths directly. We showed in Section \ref{sec:properties} that the elastic depths decrease monotonically from their unique center, as trajectories become more outlying, so using the depths directly does not incur a loss of outlyingness information. Additionally, unlike methods that place bounds on the observed data, using a boxplot on the depth values circumvents the problem of shape outliers being masked due to scale, translation, and phase variability. This is because the boundaries of a depth boxplot correspond to entire central regions on the shape space of functions, and not merely central regions on the projections of functions onto $\R$ (or $\R^2$ or $\mathbbm{S}^2$). 

\begin{figure}[H]
    \begin{center}
    \includegraphics[width=0.65\textwidth,valign=c]{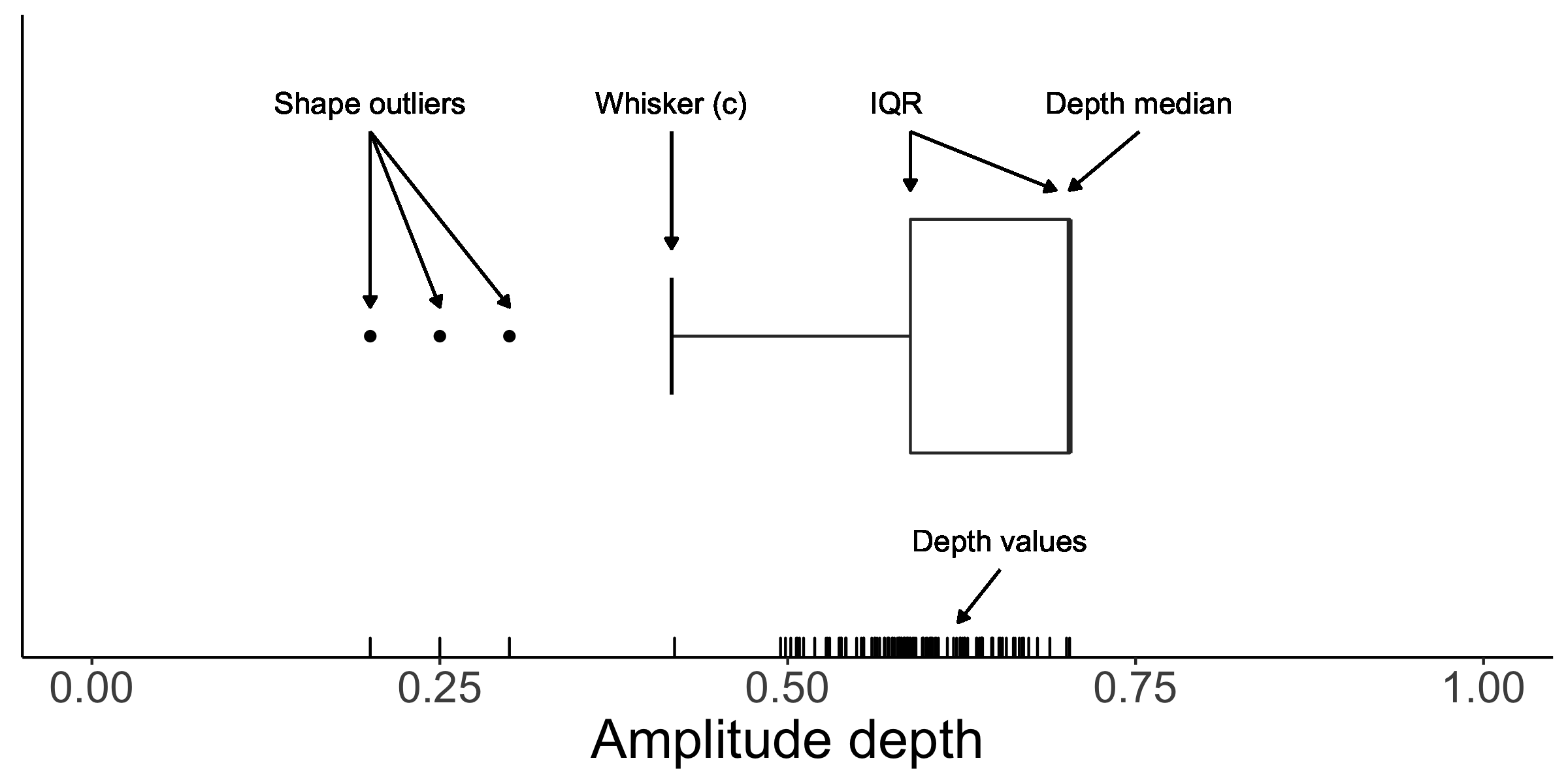}
     \caption{Diagram of the amplitude depth boxplot, created using Algorithm \ref{alg:boxplot} with $k = 1.5$, on example data. The depth median, IQR boundaries, whisker (c), and three shape outliers have all been labeled accordingly. Each of the trajectories' amplitude depths has been plotted along the horizontal axis.}
    \label{fig:diagram}
    \end{center}
\end{figure}

Algorithm \ref{alg:boxplot} describes how to construct the amplitude depth boxplot and how amplitude anomalies are identified with the whisker $c$. Phase anomalies can similarly be defined by substituting amplitude depths for phase depths.
\begin{algorithm}
    \SetKwInOut{Input}{Input}
    \SetKwInOut{Output}{Output}

    \Input{Functions $f_1$,...,$f_n$ and multiplier $k$}
    \Output{Outlier status of $f_1$,...,$f_n$ given $k$}
    
    \For{$i\gets1$ \KwTo $n$}{
    Compute amplitude depths $D_{A, n}(f_i, P)$
    }
    Compute $IQR = $ $\text{max}\{D_{A, n}(f_i, P)\}$ $ - \text{median}\{D_{A, n}(f_i, P)\}$ \\
    Compute $c = \text{median}\{D_{A, n}(f_i, P)\} - k \times IQR$.
    
    \For{$i\gets1$ \KwTo $n$}{
      \eIf{$D_{A, n}(f_i) < c$} {
        $f_i$ is an outlier
      } { $f_i$ is not an outlier
      }
    }
    \caption{Depth boxplots for finding amplitude outliers}
    \label{alg:boxplot}
\end{algorithm}
The boxplot created in Algorithm \ref{alg:boxplot} consists of the following three pieces: The median, the IQR, and the whisker $c$ (Figure \ref{fig:diagram}). The median of the boxplot is the largest depth, because as was shown in Section \ref{sec:properties}, the largest depth corresponds to the median of the distribution. The IQR is the 50\% central region because, as per the IQR of univariate data, this range contains the inner 50\% of the data. Most importantly is the whisker value $c$, which determines which trajectories are considered outliers. Any trajectory with an amplitude depth of less than $c$ is considered an anomaly because it is statistically too far from the rest of the data. 

The whisker $c$ is determined by a multiplier or inflation factor $k$. The quantity $k$ is a free parameter that must be set to detect anomalies. In classical univariate boxplots, $k = 1.5$, so as to achieve approximately 99.3\% coverage of the boxplot on Gaussian data. This guarantee does not necessarily extend to functional data, but we find empirically that $k \in [1.5, 2]$ works well to separate outliers from inliers as long as the sampling frequency is high enough to fully represent the functional data. We investigate the depth boxplot's dependency on $k$ numerically in Section \ref{sec:sensitivity} of the online supplement and find that detection performance is fairly robust to $k$ for a wide range of values and across many types of data. More details on the choice of $k$ can also be found in Section \ref{sec:multi}.

\subsection{Depth Thresholding} \label{thresholding}
Boxplots and other hard cutoffs are not the only way to investigate shape anomalies. The quantiles of the depth distribution itself can also be used to investigate the most extreme data. The elastic depths induce a proper center-outward ordering of the trajectories, so the most extreme, i.e. smallest, depth values correspond to the most extreme trajectories. Therefore, if we wanted to view the 5\% most extreme functions, we could simply select the functions with the 5\% smallest elastic depth values. More generally, the $100 \times p$\% most outlying functions have depth values below the $(1-p)^{th}$ quantile of the depth's distribution. This type of thresholding is quite useful in exploratory analysis for comparing, say, the 1\%, 5\%, and 10\% most outlying shapes with the 1\%, 5\%, and 10\% most inlying shapes.

The limitation of thresholding is that it will always select $p$\% of the data to be outlying, so as an anomaly detector it is insufficient on its own. However, it can be paired with the depth boxplots to produce a more robust depth boxplot. Algorithm \ref{alg:thresh_boxplot} extends algorithm \ref{alg:boxplot} to include a thresholding parameter $p$ so that to be considered an outlier, a function must have a depth value below the whisker $c$ and below the $(1-p)^{th}$ quantile of the depth's distribution. 
\begin{algorithm}
    \SetKwInOut{Input}{Input}
    \SetKwInOut{Output}{Output}

    \Input{Functions $f_1$,...,$f_n$, multiplier $k$, and threshold $p$}
    \Output{Outlier status of $f_1$,...,$f_n$ given $k$ and $p$}
    
    \For{$i\gets1$ \KwTo $n$}{
    Compute amplitude depths $D_A(f_i, P)$
    }
    Compute $IQR = $ $\text{max}\{D_A(f_i, P)\}$ $ - \text{median}\{D_A(f_i, P)\}$ \\
    Compute $c = \text{median}\{D_A(f_i, P)\} - k \times IQR$. \\
    Compute $q = (1-p)^{th}$ quantile of $\{D_A(f_i, P)\}$
    
    \For{$i\gets1$ \KwTo $n$}{
      \eIf{$D_A(f_i) < \min\{c, q\}$} {
        $f_i$ is an outlier
      } { $f_i$ is not an outlier
      }
    }
    \caption{Depth boxplots for finding amplitude outliers with thresholding}
    \label{alg:thresh_boxplot}
\end{algorithm}

The purpose of $p$ is mainly to control the number of false positives when detecting outliers. If $p = 0.95$, then at most 5\% of the trajectories will be considered outlying, no matter what the whisker value is. While the whisker is generally sufficient for achieving good coverage of the depth distribution and detecting anomalies, there are situations, such as low sampling frequency (Section \ref{sec:sensitivity} of the online supplement), where the whisker can fall short unless the multiplier $k$ is set higher. In these situations, $p$ will act to effectively increase the $k$ so that better coverage is achieved. 

\section{Simulation Study} \label{simulation}

A simulation study was conducted to comprehensively assess the performance of the elastic depths and the depth boxplots. We compared our method against nine other shape anomaly detectors: the Outliergram (OG) \citep{arribasgil2014}, Sequential Transformations (ST-T1, ST-T2, ST-D1) \citep{dai2018functional}, the Functional Outlier Map (FOM) \citep{rousseeuw2018}, Total Variation Depth (TVD) \citep{huang2016total}, the Magnitude-Shape (MS) plot \citep{dai2018multivariate}, the Robust Functional Tangential Angle Pseudo-depth (rFUNTA) \citep{kuhnt2016}, Order Extended Integrated Depth (FDJ and IDJ) \citep{nagy2017}, Geometric boxplots (GEOM) \citep{xie2017geometric}, and Directional Outlyingness (DIR) \citep{dai2019}. In the following sections comparisons to TVD, MS, DIR, and GEOM are included while the rest are deferred to the appendix. These four methods were consistently the strongest competitors across each of the outlier models. We describe their implementation here briefly.

The TVD outliers were found using the \texttt{detectOutlier} function in the \texttt{TVD} R package with an empFactor = 1.5. MS outliers were found by computing the \textbf{MO} and \textbf{VO} quantities then using the \texttt{cerioli2010.irmcd.test} function in the  \texttt{CerioliOutlierDetection} R package to compute the boundary with a coverage probability of 99.3\%. DIR outliers were found using the authors \texttt{dir.out} function with Mahalanobis distance and the default parameters fac $ = 0.154$, and cutoff $= 6.91$. GEOM outliers were found using the \texttt{AmplitudeBoxplot} function in the \texttt{fdasrvf} R package using $k = 1$. Implementation details for the detectors that are deferred to the appendix have likewise been deferred to the appendix (Section \ref{sec:full_sims} of the online supplement). Elastic depths (ED) outliers were identified using the depth boxplots with $k = 1.8$. Boxplots were computed on both the amplitude and phase depths separately. The results using amplitude depth are denoted as ED-A and the results using phase depths are ED-P.

\subsection{Simulation Design}

We define seven different shape outlyingness scenarios to test the effectiveness of the above shape outlier detectors. Each of these scenarios is represented by one of the seven models detailed below. The first six correspond to amplitude (shape) outliers while the seventh is for phase outliers. 

\begin{enumerate}
    \item \textbf{Model 1 (Amplitude Increase):} Main model: $X(t) = \sin(5 \pi t) + 4t + e(t) + \delta$ and Contamination model: $X(t) = 4\sin(5 \pi t) + 4t + e(t) + \delta$, where $t \in [0, 1]$, $e(t)$ is a centered Gaussian process with covariance function $\gamma(x, x') = \exp\{-(x - x')^2/0.5 \}$, and $\delta \sim N(0, 1)$ is a random additive translation term. The purpose of $\delta$ is to shift each curve by a random amount so as to mask shape outliers that could accidentally be identified as magnitude outliers.
    
    \item \textbf{Model 2 (Amplitude Decrease):}  Main model: $X(t) = \sin(5 \pi t) + 4t + e(t) + \delta$ and Contamination model: $X(t) = \frac{1}{6}\sin(5 \pi t) + 4t + e(t) + \delta$, where $t \in [0, 1]$, and $e(t)$ is the Gaussian process from Model 1. 
    
    \item \textbf{Model 3 (Mixed Polynomials):} Main model: $X(t) = t^3 - 2t^2 + 0.5t + e(t)$ and Contamination model: $X(t) = 2t^3 + t^2 - 0.5t + e(t)$.
    
    \item \textbf{Model 4 (Covariance change):} Main model: $X(t) = \sin(5 \pi t) + 4t + e_1(t) + \delta$  and Contamination model: $X(t) = \sin(5 \pi t) + 4t + e_2(t) + \delta$, where $t \in [0, 1]$ and $e_1(t)$ and $e_2(t)$ are centered Gaussian processes with covariance functions $\gamma(x, x') = \exp\{-(x - x')^2/50 \}$ and $\gamma(x, x') = \exp\{-(x - x')^2/2 \}$, respectively.
    
    \item \textbf{Model 5 (Frequency Increase):} Main model: $X(t) = \sin(2 \pi t) + 4t + e_1(t) + \delta$  and Contamination model: $X(t) = \sin(12 \pi t) + 4t + e(t) + \delta$ where $t \in [0, 1]$ and $e(t)$ is the Gaussian process from Model 1. 
    
    \item \textbf{Model 6 (Jump contamination):} Main model: $X(t) = \sin(5 \pi t) + 4t + e(t) + \delta$ and Contamination model: $X(t) = \sin(5 \pi t) -2\mathbbm{1}_{(t < T)} + 3\mathbbm{1}_{(T \leq t)} + 4t + e(t) + \delta$, where $t \in [0, 1]$ and T is distributed uniformly on $[0.4, 0.6]$.
    
    \item \textbf{Model 7 (Phase Contamination):} Main model: $X(t) = \sin(5 \pi t) + 4t + e(t) + \delta$ and Contamination model: $X(t) = \sin(5 \pi \g(t)) + 4\g(t) + e(\g(t)) + \delta$, where $t \in [0, 1]$ and $\g$ is a random phase function from $\Gamma$. The functions $\g$ are generated from the first two Fourier basis functions with random amplitudes distributed as $N(0, \sigma)$ on the tangent space to the unit Hilbert sphere. We use $\sigma = 6$ to impose a large amount of phase variability on the contamination model (Figure \ref{fig:phase_comp}).
\end{enumerate}

\begin{figure}
\centering
  \includegraphics[width=\textwidth,valign=c]{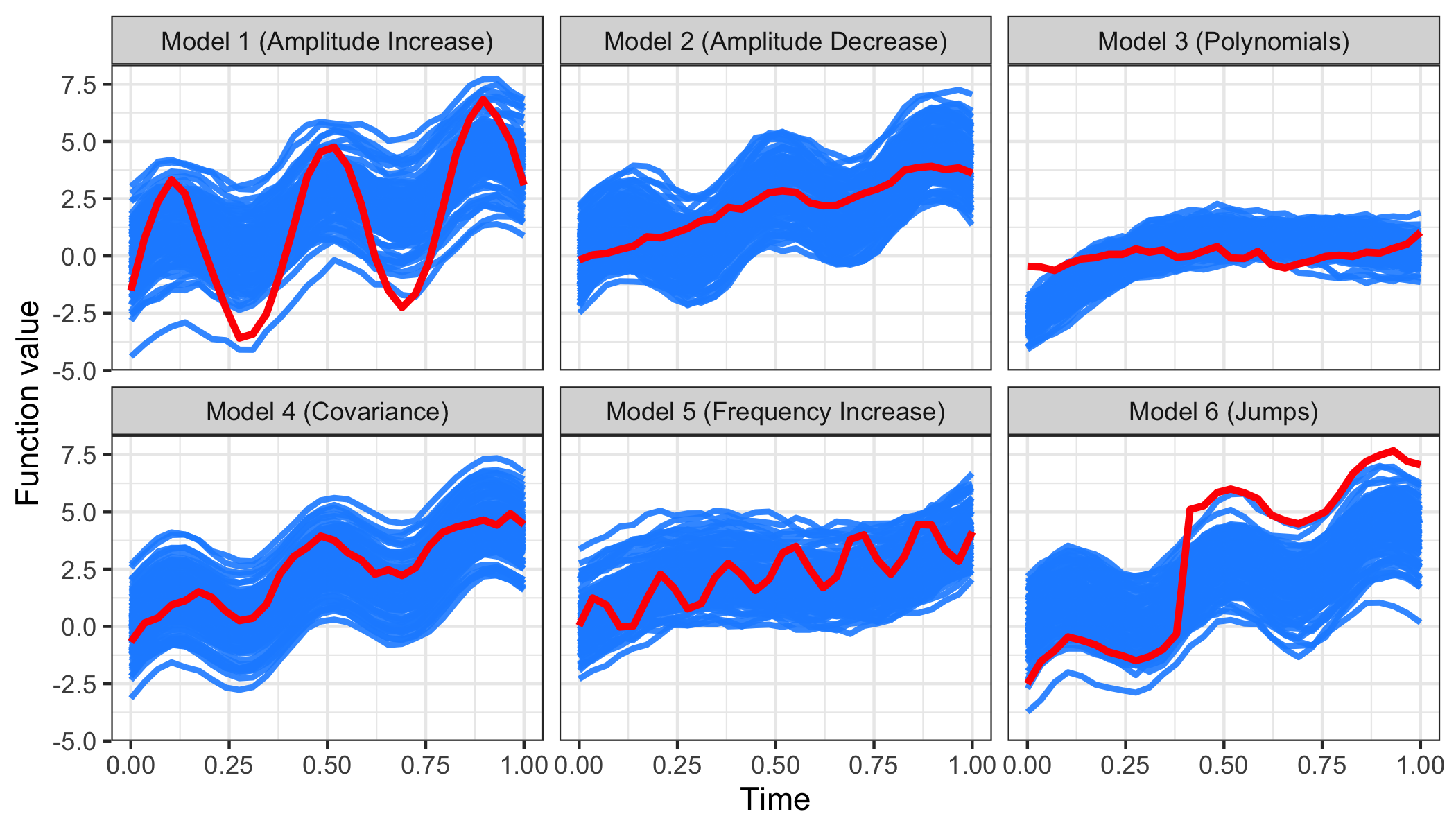}
  \caption{Main model (blue solid lines) v.s. Contamination model (red lines) in each of the amplitude outlier models.}
 \label{fig:modelexample}
\end{figure}

Each of these models, except model 7, was then further contaminated with two additional sources of noise: compositional (phase) noise and magnitude outliers. Compositional noise was added by composing each trajectory with a random phase function generated by the \texttt{rgam} function in the \texttt{fdasrvf} package with sigma = 0.1. Magnitude outliers were added by randomly shifting $10\%$ of the generated functions by $\pm 10$. 
Model 7 was only contaminated with magnitude outliers because adding phase noise would destroy the difference in phases that we are trying to detect. These two noise sources introduce a level of realism to our simulations because nuisance phase and magnitude outlyingness are often present when analyzing shapes. The base amplitude outlier models without additional phase and magnitude noise are pictured in Figure \ref{fig:modelexample}.

\subsection{Contamination by Multiple Anomalies} \label{sec:multi}
We considered the case when 10\% of the data is outlying in shape. We compared the performance of the detection methods on the seven outlier models using the $F_1$ score \citep{chinchor1992muc} for outlier classification. The $F_1$ score is a comprehensive measure of classification accuracy that considers both the precision (positive predictive value) and the recall (true positive rate) of a detection method. A method that perfectly classifies all outliers as outliers and all inliers as inliers will have an $F_1$ score of 1. Methods that do not perfectly classify will have $F_1$ scores less than 1. The $F_1$ score was traditionally defined as the harmonic mean of precision and recall, but it can also be expressed in terms of the more familiar True Positive (TP), False Negative (FN), and False Positives (FP) quantities:
\begin{align*}
    F_1 &= \frac{2TP}{2TP + FN + FP}.
\end{align*}

90 inlying trajectories and 10 outlying trajectories were sampled from the main model and contamination model respectively. Compositional noise and magnitude outliers were again added to each of the models, except for model 7 (Phase Contamination) where only magnitude outliers were added. Trajectories were sampled on an equidistant 30 point grid over $[0, 1]$ and 1000 simulations were performed for each of the six models. The results for the top models are summarized in Figure \ref{fig:comparison}. Full results for all considered models are available in the appendix. 

\begin{figure}
\centering
  \includegraphics[width=0.85\textwidth,valign=c]{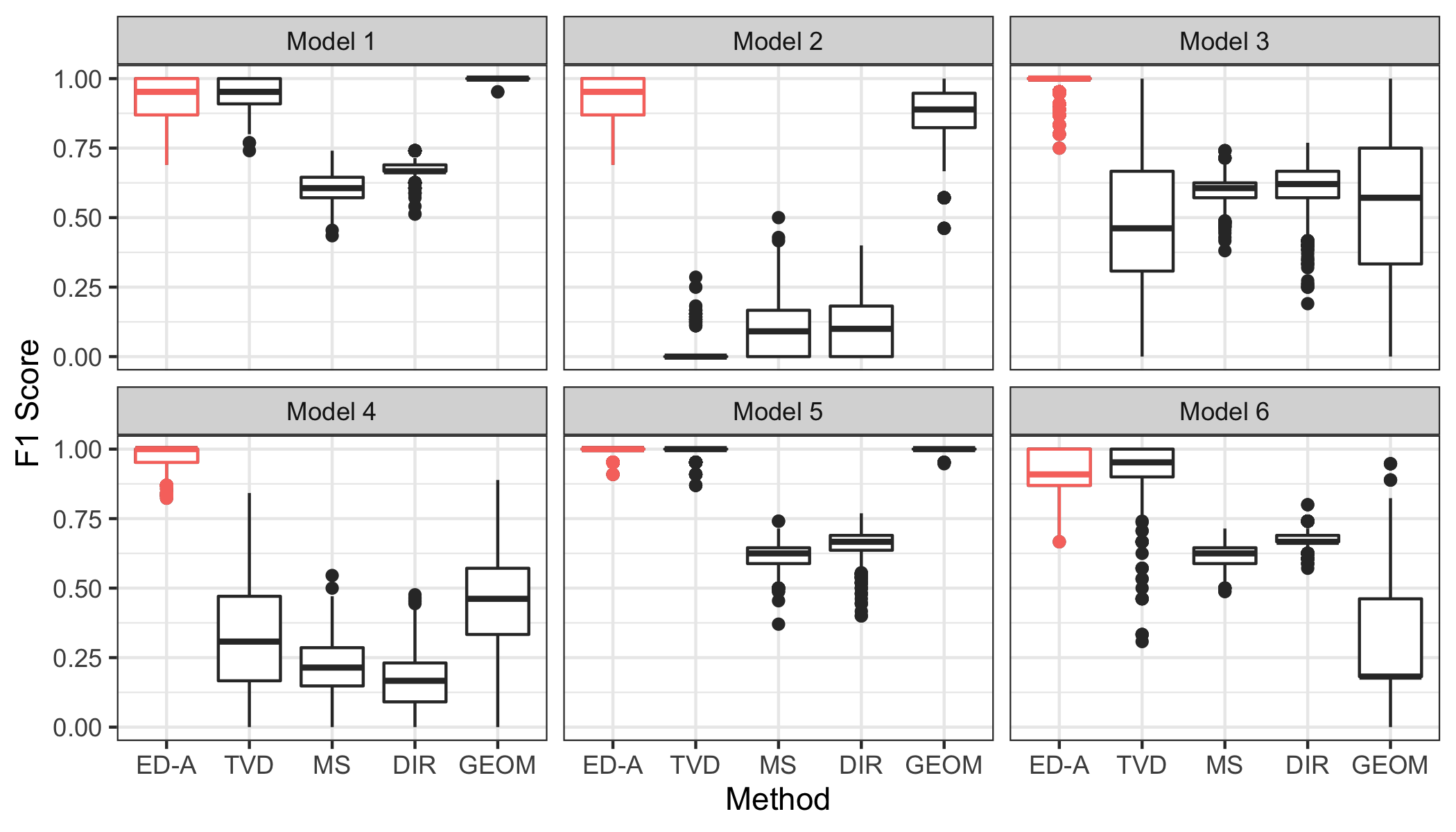}
  \caption{F1 score comparison of the top five models (ED-A, MS, DIR, TVD, and GEOM) on each of the six amplitude outlier models.}
 \label{fig:comparison}
\end{figure}

Figure \ref{fig:comparison} shows that, across the six amplitude outlier models, the elastic depths maintained the highest average $F_1$ score. The amplitude depth based boxplots have an average $F_1$ score of around 0.95 - 1.0, indicating that in each scenario, regardless of outlier or inlier type, the amplitude depths can achieve near-perfect detection. Consistently high performance of the elastic depths is notable because existing methods, while strong in some cases, suffer major losses of power in others. For instance, on some models, such as model 3 (mixed polynomials) and model 4 (covariance change), the amplitude depth based boxplot was the only method able to detect the shape outliers consistently. Even GEOM, which uses the elastic distances, was unable to consistently separate these outliers from the inliers. Together, the results show that because the elastic depth based boxplots use the shape distribution, albeit indirectly via data depth, they can consistently and skillfully detect a wide variety of shape outliers. They do not generally suffer a loss of power due to compositional (phase) noise, translation noise, presence of magnitude outliers, or even inlier and outlier type.

\begin{figure}[H]
    \begin{center}
    \includegraphics[width=0.8\textwidth,valign=c]{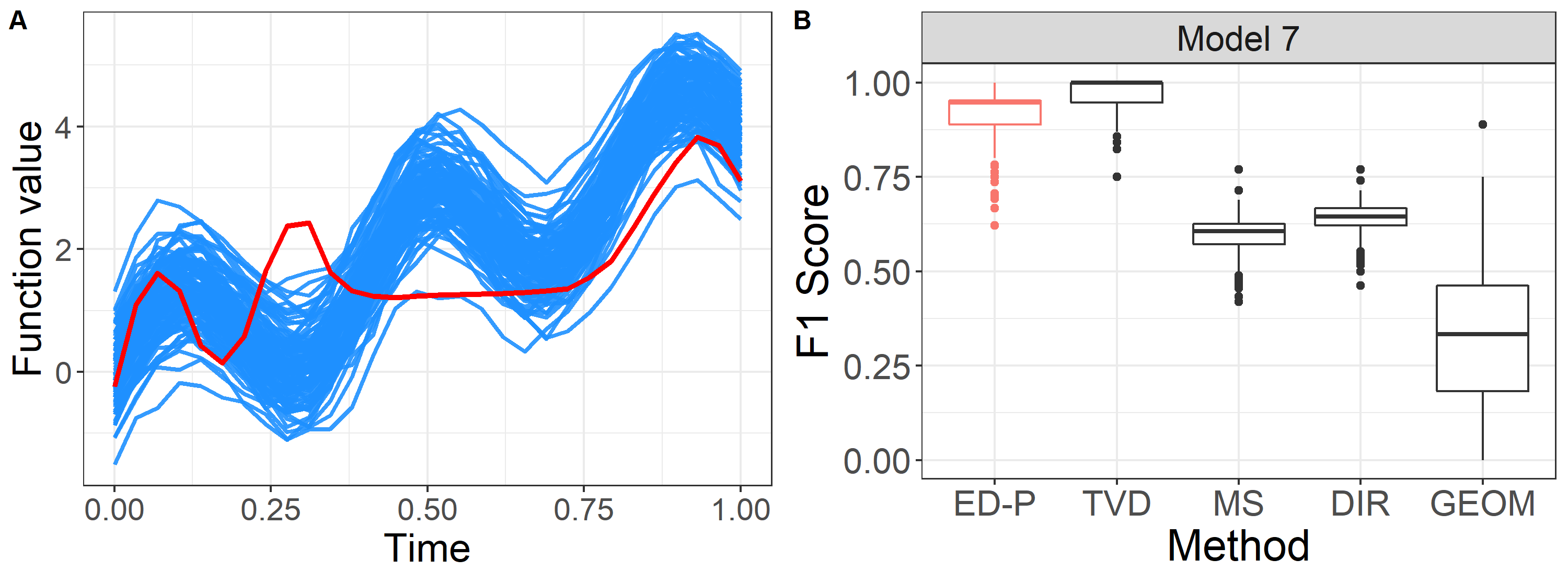}
  \caption{Panel A: Example inlier trajectories (blue) v.s. a phase outlying trajectory (red) Panel B: $F_1$ score comparison of the top five models (ED-P, MS, DIR, TVD, and GEOM) on the phase outlier model. ED-P and TVD both have nearly perfect detection rates, with TVD slightly winning out in this simulation.}
  \label{fig:phase_comp}
    \end{center}
\end{figure}

Figure \ref{fig:phase_comp} shows that, across the top phase outlier detection models, ED-P, the phase counterpart of ED-A, has near-perfect detection skill. The TVD, however, maintains a slightly higher $F_1$ score. Our conjecture is that the data generation process may have incidentally induced non-phase based differences into the phase outliers.  TVD, MS, and DIR could potentially take advantage of this non-phase information to improve their score, whereas ED-P and GEOM could not because they only use phase information. See a small simulation example in Section A.6 in the online supplement.

We also investigated the sensitivity of the boxplots to the parameter $k$ via a simulation study in Section \ref{sec:sensitivity} of the online supplement. Overall, we found that the coverage of the boxplot on the inlying uncontaminated data is insensitive to the value of $k$. Across each model, the coverage of the boxplots steadily increases from about 95\% at $k = 1$ to around 100\% at $k = 3$. Some models required higher values of $k$ to achieve the desired 99\% coverage, which we found to be due to under sampling, i.e. sampling below the Nyquist rate, of the observed trajectories. We recommend avoiding this potential issue by ensuring adequate sampling of the trajectories when possible or using $k \approx 2$ when this is not possible.

\section{Hurricane Trajectories} \label{sec:hurricanes}
Our motivating example (Figure \ref{fig:hurricanes}) comes from the National Hurricane Center's (NHC) Atlantic Hurricane Database (HURDAT2) \citep{landsea2013atlantic}. The NHC assimilates all observations, real time and post-storm, for each tropical cyclone to estimate and record its characteristics and path across the Atlantic Ocean. The HURDAT2 database contains records for 979 tropical cyclone paths of various lengths, shapes, sizes, orientations, and placements. 
We only consider storms with at least 25 observations because only those paths had sufficient time to develop. The storms were then further subset to include only those originating in the ocean and north of South America.

\begin{figure}
\centering
  \includegraphics[width=0.7\textwidth,valign=c]{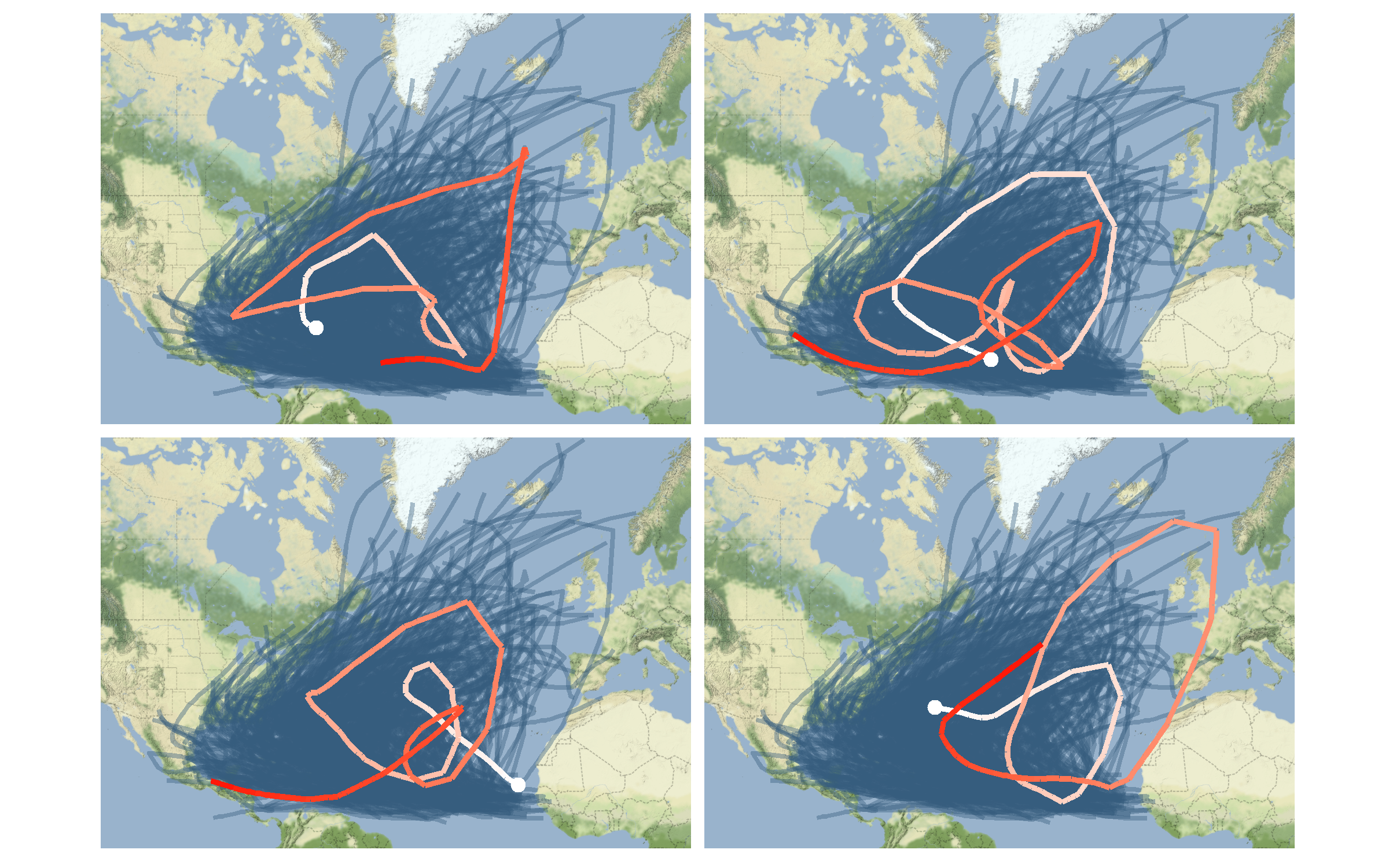}
  \caption{Four most shape outlying hurricane trajectories from the HURDAT2 data overlaid on the entire data set. The starting point for each track is marked by a point and trajectories become progressively darker as they develop.}
  \label{fig:hurr}
\end{figure}

The typical path of a hurricane is ``U'' shaped, starting in Africa then cutting across the eastern United States and finally heading back east towards Europe.
Due to the vast distances hurricanes travel it would be inappropriate to treat them as lying on a Euclidean plane. Instead, we consider them as trajectories on the surface of a unit sphere $\mathbb{S}^2$. We used the elastic depth boxplots with a depth threshold of 0.05 to limit the number of amplitude outliers to fourteen; the top four of which are pictured in Figure \ref{fig:hurr}. 

Each of the top four outliers is markedly different from the standard ``U'' shape. They exhibit an atypical spiraling behavior as they meander across the Atlantic. The identification of shape outliers helps climate scientists further investigate what causes the trajectories to be anomalous. It can be very important for improving the accuracy of hurricane prediction algorithms if the dynamics which produce anomalies are well understood. Further data examples on $\R$ and $\R2$ valued trajectories can be found in section \ref{sec:examples_appendix} of the online supplement.

\section{Discussion} \label{discussion}

In this paper, we proposed a new class of functional depths based on the elastic distance metrics and showed how they may be used to detect shape outliers. The theoretical properties of our new elastic depth were investigated, and it was shown that they satisfy most key properties required of a depth metric. These include translation, scale, rearrangement, phase invariance (equivariance), maximality of the center, convex level sets and monotonicity from the center. Rearrangement invariance and phase invariance were particularly crucial for detecting shape anomalies because they allowed the amplitude depth to measure centrality independent of phase.

We demonstrated the empirical performance of our method together with nine competing methods using extensive simulation studies. It was shown that our method attains the overall highest average $F_1$ score across all models. On each of the six amplitude models the amplitude depth based boxplots had an average $F_1$ score of around 0.95 - 1.0, indicating that in each scenario, regardless of outlier or inlier type, the amplitude depths achieved near-perfect detection on average. On some models, such as model 3 (mixed polynomials) and model 4 (covariance change), the amplitude depth based boxplot was the only method able to detect the shape outliers consistently. Together, these results demonstrate the power of using the shape distribution, albeit indirectly via data depth, to detect shape outliers.

The simulation results, of course, depend on the boxplot multiplier $k$. We recommend setting $k = 2$ as the default value. We found through empirical studies (See online supplement Section \ref{sec:sensitivity}) that any $k$ value between 1.5 and 2.25 provides nearly the same level of detection skill. The value $k = 2$ had the most favorable trade-off between false positives and false negatives across all of the outlier models in the simulation. Larger values (up to around 2.5-2.8) are helpful when the sampling frequency of the trajectories is too low (See online supplement Section \ref{sec:sensitivity}). If the sample size is large then the depth quantiles, $p$, can also be used to help set a minimum value for $k$. That is, choose $k$ large enough to make the whisker value less than the $(1-p)$'th quantile of the amplitude depth. This guarantees that the shapes of the inner $p \times 100$\% of trajectories will be considered as inliers.

Finally, we showed how the elastic depths may be used to identify shape outliers in functional data observed on the unit sphere $\mathbb{S}^2$. We used the HURDAT2 hurricane track database and identified the four most shape outlying trajectories. We found that these trajectories' paths were remarkably different from the standard U-shaped paths that hurricanes normally follow. Further applications in the online supplement \ref{sec:examples_appendix} demonstrate the elastic depths on $\R$, $\R^2$ valued functional data. These data examples illustrated the simplicity and consistency with which the elastic depths may be applied, regardless of the underlying geometry of the space.


\section*{Acknowledgments}
This paper describes objective technical results and analysis. Any subjective views or opinions that might be expressed in the paper do not necessarily represent the views of the U.S. Department of Energy or the United States Government.
This work was supported by the Laboratory Directed Research and Development program at
Sandia National Laboratories, a multi-mission laboratory managed and
operated by National Technology and Engineering Solutions of Sandia,
LLC, a wholly owned subsidiary of Honeywell International, Inc., for the
U.S. Department of Energy's National Nuclear Security Administration
under contract DE-NA0003525.

\section*{Supplementary Materials}
\begin{description}
\item[Supplementary results:] PDF file ``Supplementary results" containing detailed descriptions of the elastic distances, proofs for all results in Section \ref{sec:properties}, additional simulations, and two additional real data examples. (pdf file).

\item[Source code:] Zip file ``Source code" containing code to produce all figures and results in the manuscript and supplementary file. (zip file)

\item[R-package for Elastic Depth:] R-package ``elasticdepth” containing code to compute the elastic depths and depth boxplots for $\mathbb{R}$, $\mathbb{R}^2$, and $\mathbb{S}^2$ valued trajectories. Available through GitHub:
\texttt{trevor-harris/elasticdepth}. (GNU zipped tar file)
\end{description}

\bibliography{refs}

\appendix
\section{Appendix}

\subsection{Amplitude distance} \label{sec:elastic_definitions}
In this section, we will introduce the explicit definitions and mathematical motivation for the amplitude distances mentioned in Section 2.2. Later parts of this section rely on concepts from topology and differential geometry such as tangent spaces and parallel transport. A thorough review of these topics, as applied to the problem of shape analysis, can be found in Chapter 3 of \citet{srivastava2016functional}. Here we will only provide conceptual introductions as they are needed to define amplitude metrics.

\subsubsection*{Phase Space} \label{sec:phase}
To define the concept of amplitude we first need to define the concept of phase, because amplitude is the properties of a function that are independent of phase. The phase space of the unit interval $[0, 1]$ is defined as 
\[
\Gamma = \{\g : [0, 1] \mapsto [0, 1] \ | \ \g(0) = 0 \text{, } \g(1) = 1 \text{, } \g \text{ is diffeomorphic}\}.
\]
This diffeomorphic constraint gives rise to the notion of elasticity because the elements of $\Gamma$, i.e. phase functions, can only smoothly stretch and contract portions of the unit interval so that it maps back to itself. Phase is generally thought of as the representation of a trajectory because any trajectory with domain $[0, 1]$ can be \textit{warped} by a phase function to appear differently. The amplitude will be taken to be those features of a trajectory that remain unchanged under any possible warping. In the following sections, we will consider functions with domains on $[0, 1]$ which take values in $\mathbb{R}^2$ and the unit sphere $\mathbb{S}^2$ and develop proper distance metrics on their amplitudes.

\subsubsection*{Amplitude distance for $\R^n$ valued functions} \label{sec:rn}
Let $F_{\mathbbm{R}^n} = \{f:[0, 1] \mapsto \mathbb{R}^n, f \text{ differentiable} \}$ be the class of differentiable trajectories on $[0, 1]$ mapping to $\R^n$ with $n \geq 2$. In higher dimensional Euclidean space ($n \geq 2)$, the scale, rotation, and phase of a trajectory have to be accounted for in order to isolate its shape. Scale variability is removed by standardizing each trajectory to have length one, that is each trajectory is divided by the $L^2$ norm of its gradient:
\[
f(t) \mapsto f(t) / ||f'|| \quad \forall t \in [0, 1],
\]
where $||f'|| = \sqrt{\int_0^1 f'(t)^2dt}$. This operation can intuitively be thought of as standardizing the ``zoom level'' or ``magnification'' of each trajectory. In Section 2.3, the $\mathbbm{R}$ valued trajectories were not length standardized because, on $\mathbbm{R}$, scale and amplitude are essentially the same after trajectories are aligned.

Accounting for rotation variability requires introducing the space of rotation matrices, the Special Orthogonal Group $SO(n)$, and another level of optimization. The space $SO(n)$ is defined as the group of orthogonal matrices with determinant one. For trajectories in $\mathbbm{R}^2$, we use $SO(2)$ to define rotations around a point, in $\mathbbm{R}^3$ the group $SO(3)$ defines rotations around a line, etc. The action of $SO(n)$ on a trajectory $f$ is denoted as $O(f)$ and is defined pointwise as 
\[ O(f) = \{O f(t) : \forall t \in [0, 1] \},\]
where $O f(t)$ represents standard matrix multiplication of the $n \times n$ matrix $O$ and the $n \times 1$ vector $f(t)$. See \citet{popov1994orthogonal} for more details on the Orthogonal groups and their properties. Finding the optimal rotation matrix in $SO(n)$ is done alongside finding the optimal phase function when computing the amplitude distance.

To represent phase variability in $\mathbbm{R}^n$ valued function we can again use elements of the space $\Gamma$. This is because $\Gamma$ is only defined with respect to the domain $[0, 1]$ and not the range. As in the real valued function case, we want a distance that is invariant to simultaneous reparameterizations so we use the Square Root Velocity Function (SRVF), which was introduced in \citet{srivastava2011} as the following transformation on trajectories $f \in F_R^n$:
\begin{definition} \label{def:srvf}
Let $f$ be a differentiable trajectory in $F_R^n$, the Square Root Slope Function (SRVF) of $f$ is
\[
q_f(t) = \frac{f'(t)}{\sqrt{|f'(t)|}},
\]
where $|f'(t)|$ is the absolute value of $f'$ at $t$.
\end{definition}
This definition is effectively the multivariate generalization of Definition 2.1. The only difference is that, due to the length restriction on $f$, this transformation maps trajectories onto the $L^2$ ball of radius one instead of $L^2$ space itself \citep{srivastava2016functional}. Without this restriction the SRVF would map to $L^2$ and be an exact generalization of the SRSF.

This definition is nearly identical to that of 2.1, except that the trajectory $f$ takes values in $\mathbbm{R}^n$. Due to the length constraint on $f$, this transformation maps the length constrained functions onto the $L^2$ ball of radius one, instead of $L^2$ space itself \citep{srivastava2016functional}. This means the norm on the SRVFS is instead the arc length distance on $L^2$ spheres which is defined as
\begin{equation} \label{eqn:Rnmetric}
    d(q_f, q_g) = \arccos{{\int_0^1 \langle q_f(t), q_g(t) \rangle dt}},
\end{equation}
where $q_f = SRVF(f)$ and $q_g = SRVF(g)$ for two trajectories $f, g \in F_{R^n}$ and $\langle q_f(t), q_g(t) \rangle$ denotes the inner product of the vectors $q_f(t), q_g(t)$. To convert this into an amplitude distance we need to place $f$ and $g$ in phase and rotation with each other. This optimization is summarized as the following amplitude distance on $\mathbbm{R}^n$ valued trajectories \citep{srivastava2011}:
\begin{definition} \label{eqn:Rndist}
Let $f$ and $g$ be two trajectories in $F_{R^n}$, then the amplitude distance between $f$ and $g$ is
\[
d_a(f, g) = \inf_{\g \in \Gamma, O \in SO(n)} \arccos{{\int_0^1 \langle q_f(t), q_{O(g \circ \gamma}(t) \rangle dt}},
\]
where $q_f$ and $q_{O(g \circ \g)}$ denote the SRVF's of $f$ and $O(g \circ \g)$ respectively.
\end{definition}

\subsubsection*{Amplitude distance for $\mathbb{S}^2$ valued functions} \label{sec:s2}
Let $F_{\mathbbm{R}^2} = \{f:[0, 1] \mapsto \mathbb{S}^2, f \text{ differentiable} \}$. For differentiable functions constrained to live on a sphere, rotation about a point and scaling are not possible. However, the direct comparison of function gradients (SRSVs) is also not possible without additional steps \citep{srivastava2016functional}. The gradient of a function at a particular point lives in a vector space perpendicular to $\mathbbm{S}^2$ known as the tangent space. The tangent space represents all possible ``directions'' in which a trajectory can pass through a given point, and can be viewed as a flat plane touching $\mathbbm{S}^2$ only at that point. 

On a nonlinear manifold, such as $\mathbbm{S}^2$ the tangent space changes depending on which point on the manifold it touches. This means the distance between gradients (SRVFs) is not well defined because they don't live in the same space. To remedy this issue \citet{su2014statistical} introduced the Transported Square Root Velocity Function (TSRVF), which uses the idea of parallel transport from differential geometry to make the tangent spaces of two trajectories on $\mathbbm{S}^2$ comparable. Roughly speaking, parallel transport describes how to transform the tangent space of one location into the tangent space of another location, provided the two locations can be connected by a smooth path \citep{kobayashi1996wiley}. On the unit sphere $\mathbbm{S}^2$ parallel transport is a straightforward analytic calculation \citep{srivastava2016functional}.
\begin{definition} \label{def:transport}
    Let $f$ be a differentiable trajectory in $F_{\mathbb{S}^2}$ and let $c \in \mathbb{S}^2$. Then $\forall t \in [0, 1]$ the parallel transport of $f'(t)$ to the tangent space of the point $c$, along the shortest available path, is defined as
    \[
    f'(t)_{f(t) \mapsto c} =  f'(t) - 2 \langle f'(t), c \rangle \frac{f(t) + c}{||f(t) + c||^2}.
    \]
\end{definition}
This definition means that $f'(t)$, which is tangent to $f(t)$, can be converted into tangent vector at any point $c$ on the sphere. More importantly this means that the entire gradients of two trajectories $f, g$ can both be moved into the tangent space of a single point. Once both trajectories completely share a common tangent space, the amplitude distance between $f, g$ is computable. This leads to the notion of the TSRVF in \citep{su2014statistical}:
\begin{definition} \label{def:tsrvf}
Let $f$ be a differentiable trajectory in $F_M$ for some Riemannian manifold $M$ with norm $||\cdot||$. The Transported Square Root Vector Field (TSRVF) of $f$ at the point $f(t)$ to the point $c \in M$ is
\begin{equation}
    h_f(t) = \frac{f'(t)_{f(t) \mapsto c}}{\sqrt{|f'(t)|}},
\end{equation}{}
Where $f'(t)_{f(t) \mapsto c}$ represents the parallel transport of the tangent vector $f'(t)$ from $f(t)$ to $c$.
\end{definition}

To define an amplitude distance, a common point $c \in \mathbbm{S}^2$ must be chosen to transport each function to. As \citet{su2014statistical} showed, the choice of $c$ only minorly impacts the computed distances and does so in a consistent way. We also have to account for phase variability as we did for 
$\mathbbm{R}$ and $\mathbbm{R}^n$ valued trajectories. This leads to the following definition of amplitude distance for trajectories on $\mathbbm{S}^2$ \citep{su2014statistical}:
\begin{definition} \label{eqn:S2dist}
Let $f$ and $g$ be two trajectories in $F_{S^2}$, then the amplitude distance between $f$ and $g$ is
\[
d_a(f, g) = \inf_{\g \in \Gamma}||h_f - h_{g \circ \g}||_2,,
\]
where $h_f$ and $h_{g \circ \g}$ denote the TSRVF's of $f$ and $g \circ \g$ respectively to some common point $c \in \mathbbm{S}^2$.
\end{definition}

\subsubsection*{Phase distance} \label{phase}
The phase distance is the same for each of the three manifolds and is, in fact, the same for all univariate functions mapping to any Riemannian manifold. This is because the phase space $\Gamma$ is not defined with respect to the range of the functions, $M$, but only to the domain $[0, 1]$.
To define phase distance we use the optimal $\gamma$ that defines the amplitude distance.
The phase space $\Gamma$ is a nonlinear manifold with no known geometry so we use the SRSF to map $\Gamma$ to a known geometry. Phase functions are positive for all $t \in [0, 1]$ and $||q_{\gamma}|| = \int_0^1 \sqrt{\gamma'(t)}^2 dt = 1$, so the SRSF maps $\Gamma$ onto the positive orthant of a unit Hilbert Sphere. Thus the phase distance is defined as 
\begin{equation} \label{phsdistance}
        d_{p}(f_1, f_2) = \arccos{{\int_0^1 \langle \sqrt{I'(t)}, \sqrt{\gamma'(t)} \rangle dt}},
\end{equation}
where $I(t) = t$ is the identity function and $\langle \sqrt{I'(t)}, \sqrt{\gamma'(t)} \rangle$ is the inner product between the two vectors $\sqrt{I'(t)}$ and $\sqrt{\gamma'(t)}$ \citep{srivastava2016functional}. The metric $d_p(f_1, f_2)$ is essentially measuring the amount of elastic deformation needed to compare the amplitudes of $f_1$ and $f_2$.


\subsection{Proofs} \label{sec:proofs}

\noindent \textbf{Proof of Proposition 3.1 (Phase Invariance)} 
\begin{proof}
Let $f \in F_M$ and let $\g \in \Gamma$. Then 
\begin{align*}
    D_a(f \circ \g, P) &= (1 + O(f \circ \g, P))^{-1} \\
                      &= \text{median}\{d_a(f \circ \g, X): X \in F_M\} \\
                      &= \text{median}\{d_a(f, X): X \in F_M\} \\
                      &= (1 + O(f, P))^{-1} \\
                      &= D_a(f, P) 
\end{align*}
by phase invariance of the amplitude distances \citep{kurtek2011Signal}.
\end{proof}

\noindent \textbf{Proof of Proposition 3.2 (Maximality of the Center)} 
\begin{proof}
The Fr\'echet median $m$ of the amplitudes is defined as the trajectory which minimizes the median amplitude distance between itself and all other trajectories in $F_M$ \citep{srivastava2016functional}. That is,
$m = \arg\min_{f \in F_M} \text{median}\{d_a(f, X): X \in F_M\}$.
Therefore if $s = \arg\max_{f \in F_M} D_a(f, P)$ then:
\begin{align*}
    s &= \arg\max_{f \in F_M} D_a(f, P) \\
      &= \arg\min_{f \in F_M} O_a(f, P) \\
      &= \arg\min_{f \in F_M} \text{median}\{d_a(f, X): X \in F_M\} \\
      &= m.
\end{align*}
Thus the maximizer of the amplitude depth is the Fr\'echet median of the amplitudes.
\end{proof}

\noindent \textbf{Proof of Proposition 3.3 (Convex Level Sets)} 
\begin{proof}
Let $f_1$ and $f_2$ be in $F_M$ with amplitudes $[q_1]$ and $[q_2]$ in $D_{a, \alpha}(P)$.
Let $f \in F_M$ such that the amplitude of $f$ is $[q] = \lambda[q_1] + (1-\lambda)[q_2]$ for some $\lambda \in [0, 1]$. The amplitude distance between $f$ and a random $X \sim P$ in $F_M$ with amplitude $[q_X]$ can be upper bounded as follows 
\begin{align*}
    d_a(f, X) &= d_a(\lambda[q_1] + (1-\lambda)[q_2], [q_x]) \\
    &\leq \lambda  d_a([q_1], [q_X]) +  (1 -  \lambda) d_a([q_2], [q_X]) \\
    &= \lambda  d_a(f_1, X) +  (1 - \lambda) d_a(f_2, X),
\end{align*}
by the convexity of the amplitude distance. Since this is a convex combination of positive real numbers, $d(\cdot, \cdot)$, we have that
\begin{align*}
    \text{median}(d_a(f, X)) &\leq \text{median}(\lambda d_a(f_1, X) +  (1 - \lambda) d_a(f_2, X)) \\
    &\leq \max\{\text{median}(d_a(f_1, X)), \text{median}(d_a(f_2, X))\}.
\end{align*}
Since $D_a(f_1, P) \geq \alpha$ and $D_a(f_2, P) \geq \alpha$, by virtue of $[q_1]$ and $[q_2]$ in $D_{a, \alpha}(P)$, we see that the outlyingness functions are equivalently bounded, i.e. 
\begin{align*}
    O_a(f_1, P) = \text{median}(d_a(f_1, X)) &\leq \frac{1-\alpha}{\alpha} \\
    O_a(f_2, P) = \text{median}(d_a(f_2, X)) &\leq \frac{1-\alpha}{\alpha}.
\end{align*}
Consequently the outlyingness function $O(f, P)$ is also bounded
\[
O_a(f, P) = \text{median}(d_a(f, X)) \leq \frac{1-\alpha}{\alpha},
\]
hence $D_a(f, P) \geq \alpha$ and so $[q]$ is in $D_{a, \alpha}(P)$. Thus the level sets induced on the amplitudes are convex. A similar proof shows that phase level sets are convex as well because the phase distance is convex.
\end{proof}

\noindent \textbf{Proof of Proposition 3.4 (Uniform Consistency)} 
\begin{proof}
Define the $\epsilon$-bracket $[l, u]$ as the set of all functions $f \in F_M$ such that $l < f < u$ and $|O_a(u, P) - O_a(l, P)| \leq \epsilon$. Because $F_M$ is a set of differentiable functions, only a finite number of $\epsilon$-brackets are needed to cover $F_M$. Therefore for any $f \in F_M$ there exists bracket $[u_i, l_i]$ such that
\begin{align*}
O_{a, n}(f, P) - O_a(f, P) &= (O_{a, n}(f, P) - O_a(u_i, P)) + (O_a(u_i, P) - O_a(f, P)) \\
                    &\leq (O_{a, n}(u_i, P) - O_a(u_i, P)) + \epsilon.
\end{align*}
Consequently,
\[
\sup_f\{O_{a, n}(f, P) - O_a(f, P)\} \leq \max_i(O_{a, n}(u_i, P) - O_a(u_i, P)) + \epsilon.
\]
By the strong law for sample quantiles $O_{n}(f, \mathbbm{P}) \xrightarrow{a.s.} O_a(f, \mathbbm{P})$ for any fixed $f \in F_M$ so the right hand side goes to $\epsilon$ almost surely. Therefore if we take an $\epsilon$ sequence converging to 0 we get $\sup_f\{O_{a, n}(f, P) - O_a(f, P)\} \xrightarrow{a.s.} 0$. Since $|O_{a, n}(f, P) - O_a(f, P)|$ forms an upper bound on $|D_{a, n}(f, P_n) - D_a(f, P)|$ we get the result
\[
\sup_{f \in F} |D_{a, n}(f, P_n) - D_a(f, P)| \xrightarrow{a.s.} 0.
\]
Similarly for the phase depth $D_p(\cdot, P)$,
\[
\sup_{f \in F} |D_{p, n}(f, P_n) - D_p(f, P)| \xrightarrow{a.s.} 0.
\]
\end{proof}


\subsection{Contamination by a Single Anomaly} \label{sec:single}
We first considered the case when a single outlier is present in the data. We compared each method on their ability to rank the outlier as the most outlying function. For each of the six outlier models, we sampled 99 functions from the main model and 1 function from the contamination model, with compositional noise and magnitude outliers added to both. Each trajectory was sampled on the same equidistant 30 point grid over $[0, 1]$. 1000 simulations were used for each outlier model to estimate the average ranks in Table \ref{outlierranks}.

\begin{table} 
\footnotesize
\centering
\begin{tabular}{rlrrrrrr|r}
  \hline
 & Method & Model 1 & Model 2 & Model 3 & Model 4 & Model 5 & Model 6 & Model 7 \\ 
  \hline
  1 & ED-A & \textbf{1.000} & \textbf{1.002} & \textbf{1.002} & \textbf{1.016} & \textbf{1.000} & 1.001 & 8.808 \\
  2 & ED-P & 62.832 & 1.027 & 1.449 & 2.012 & 2.831 & 11.151 & 2.689 \\ 
  3 & TVD & \textbf{1.000} & 26.341 & 4.453 & 6.228 & \textbf{1.000} & 1.226 & \textbf{2.355} \\ 
  4 & DIR & 1.024 & 40.764 & 4.678 & 33.504 & 4.380 & 1.011 & 10.082 \\ 
  5 & MS & \textbf{1.000} & 44.306 & 2.732 & 36.242 & 2.370 & \textbf{1.000} & 9.412 \\ 
  6 & GEOM & \textbf{1.000} & 99.995 & 93.343 & 30.776 & 1.002 & 7.590 & 92.303 \\ 
  \hline
\end{tabular}
\caption{Average rank of the single outlier by detection method and outlier model. Lower ranks (closer to 1) indicate the detection method ranked the outlier as more outlying. Bold font indicates the top results in each outlier model.} 
\label{outlierranks}
\end{table}

The results are sorted so that the methods with the lowest average ranking (ED-A) across all models are listed first and those with the highest are last. On each of the seven outlier models, except for model 7, ED-A had average rankings very near 1.000. This means that in most scenarios the amplitude depths correctly rank the outlier as the most outlying function. Other methods such as TVD and ST-T1 succeed on models 1, 5, and 6 where they almost always correctly rank the outlier but fail on models 2 and 4 where their average ranks are far from 1.000. 

The results on models 2 and 4 demonstrate an important feature of the elastic depths, which is not shared by other methods. Because the elastic depths are based on proper distance metrics in the amplitude and phase spaces, they can detect trajectories that lack amplitude and phase characteristics present in the rest of the data. Having either too low of an amplitude (phase) or too high of an amplitude (phase) is both indicative of shape outlyingness. Prior methods have primarily focused on trajectories with too high of an amplitude and thus struggle to adequately handle models 2 and 4.

\subsection{Full Simulations} \label{sec:full_sims}

We now describe the full simulation setup used to compare the Elastic Depths. Total Variation Depth (TVD), the magnitude-shape plot (MS), the directional outlyingness measure (DIR), and the Geometric boxplots (GEOM) were implemented as described in Section 5.1. The three sequential transformations (ST-T1, ST-T2, and ST-D1) were implemented by hand and outliers were identified using functional boxplot function \texttt{fbplot} in the \texttt{fda} package with functional directional quantiles as the underlying depth measure. The functional outlier map (FOM) was created using the \texttt{fom} function from the \texttt{mrfDepth} package using default values and the functional directional quantiles as the underlying depth measure. The Robust Functional Tangential Angle Psuedo-depth (rFUNTA) was computed using the author's bootstrapping procedure with $probs = 0.01$. The order extended Integrated and Infimal depths (FDJ and IDJ respectively) were computed using the \texttt{depthf.fd1} function with Half-Space (Tukey) depth from the \texttt{ddalpha} R package. Outliers were found using the authors described procedure with $order = 2$ and a 1.5 IQR multiplier.

\begin{table}[H]
\footnotesize
\centering
\begin{tabular}{rlrrrrrrr}
  \hline
 & Method & Model 1 & Model 2 & Model 3 & Model 4 & Model 5 & Model 6 & Model 7 \\ 
  \hline
  1 & ED-A & \textbf{1.000} & \textbf{1.002} & \textbf{1.002} & \textbf{1.016} & \textbf{1.000} & 1.001 & 8.808 \\ 
  2 & ST-T1 & \textbf{1.000} & 20.125 & 1.226 & 15.361 & 1.058 & \textbf{1.000} & 1.979 \\ 
  3 & TVD & \textbf{1.000} & 26.341 & 4.453 & 6.228 & \textbf{1.000} & 1.226 & 2.355 \\ 
  4 & ST-T2 & 1.285 & 13.582 & 1.096 & 15.700 & 1.712 & 10.909 & \textbf{1.591} \\ 
  5 & rFUNTA & 1.430 & 23.192 & 9.278 & 7.564 & \textbf{1.000} & 2.044 & 1.802 \\ 
  6 & FOM & 1.976 & 20.389 & 3.577 & 16.464 & 2.035 & 7.014 & 3.014 \\ 
  7 & ED-P & 62.832 & 1.027 & 1.449 & 2.012 & 2.831 & 11.151 & 2.689 \\ 
  8 & DIR & 1.024 & 40.764 & 4.678 & 33.504 & 4.380 & 1.011 & 10.082 \\ 
  9 & MS & \textbf{1.000} & 44.306 & 2.732 & 36.242 & 2.370 & \textbf{1.000} & 9.412 \\ 
  10 & OG & 9.887 & 35.803 & 13.180 & 29.691 & 12.083 & 10.004 & 17.344 \\ 
  11 & FDJ & 9.576 & 40.953 & 15.899 & 38.746 & 14.959 & 11.190 & 12.002 \\ 
  12 & IDJ & 33.140 & 40.763 & 41.528 & 38.871 & 40.575 & 38.972 & 42.730 \\ 
  13 & GEOM & \textbf{1.000} & 99.995 & 93.343 & 30.776 & 1.002 & 7.590 & 92.303 \\ 
  14 & ST-D1 & 51.031 & 50.828 & 51.286 & 51.779 & 52.746 & 52.656 & 53.573 \\ 
  \hline
\end{tabular}
\caption{Average rank of the single outlier by detection method and outlier model. Lower ranks (closer to 1) indicate the detection method ranked the outlier as more outlying. Bold font indicates the top results in each outlier model.} 
\label{A:tab:outlierranks}
\end{table}

\begin{figure}[H]
  \centering
  \includegraphics[width=\textwidth, height = 25em,valign=c]{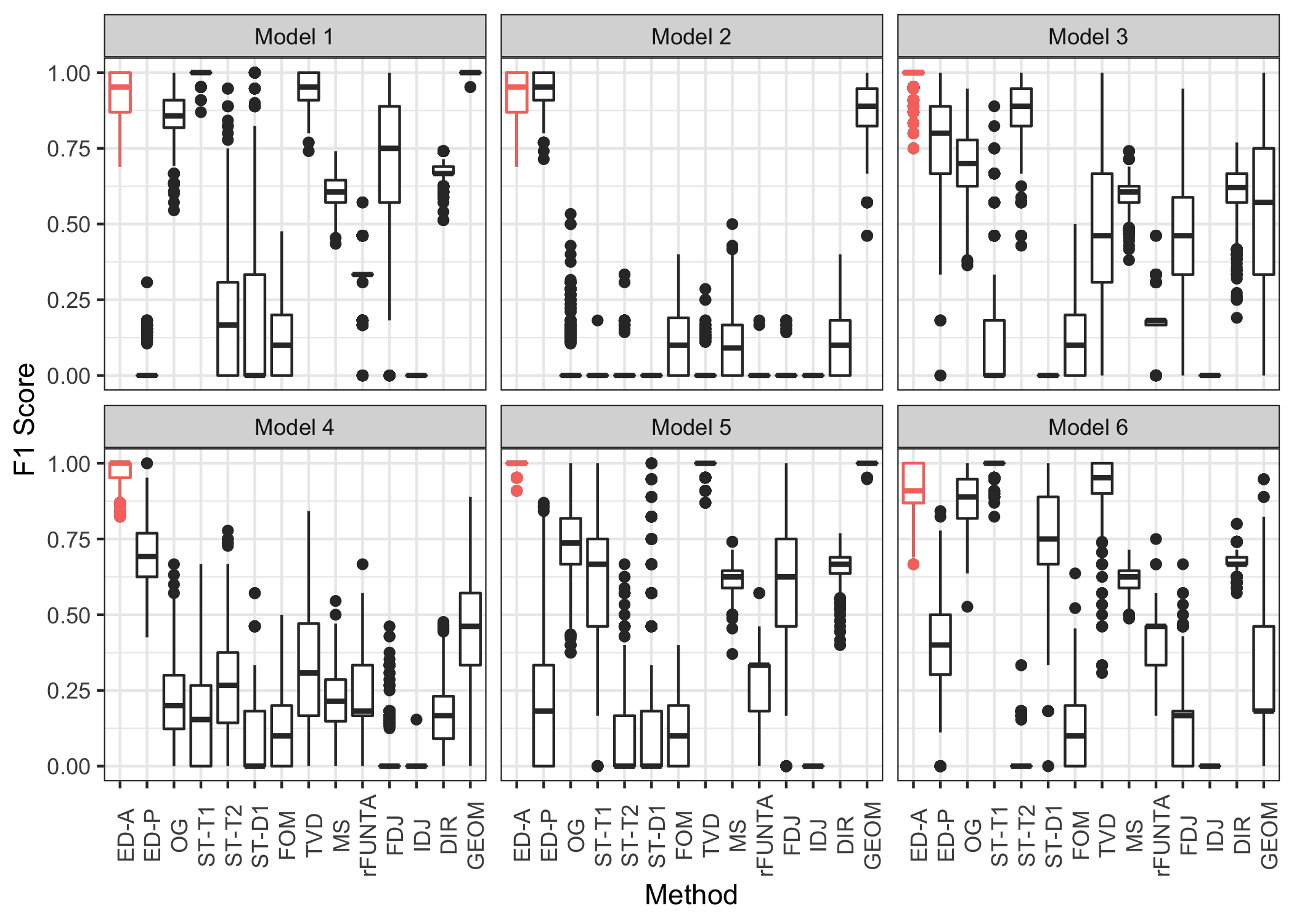}
  \caption{$F_1$ score comparison of all amplitude outlier model. ED-A, the boxplot based on amplitude depth, is the consistently highest performing method across each of the six amplitude outlier models.}
 \label{A:fig:comparison}
\end{figure}

\begin{figure}[H]
  \centering
  \includegraphics[width=0.65\textwidth,valign=c]{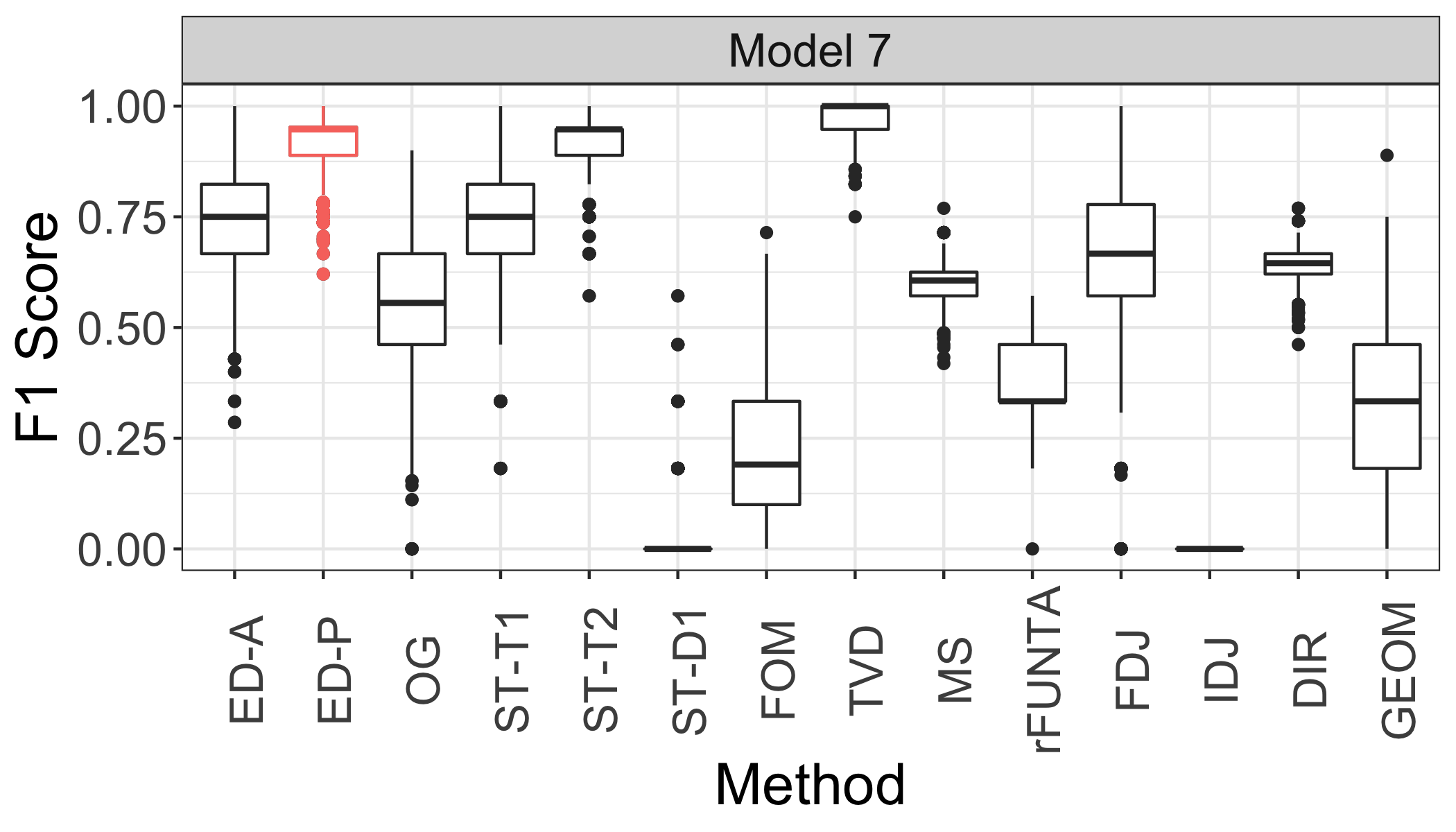}
  \caption{$F_1$ score comparison of all models. ED-A, the boxplot based on amplitude depth, is still the consistently highest performing method across each of the seven outlier models.}
 \label{A:fig:phase_comp}
\end{figure}

\subsection{Parameter sensitivity analysis} \label{sec:sensitivity}

The boxplot methodology proposed in Section 4.1 requires pre-specifying a multiplier $K$ on the IQR to determine the outlier cutoff value. In the traditional univariate case, this multiplier is set $1.5$ because, for Gaussian data, 1.5 times the IQR yields in interval covering approximately 99.3\% of the data. Theoretical justification of a similar multiplier is much more difficult in the functional data case, so in this section we numerically investigate the robustness of our outlier detection method to different multipliers.  

We first looked at the true positive rate of the elastic depths as a function of $k$. We have seen that any $k$ value less than around 2.25 (rightmost vertical black line) will have a nearly 100\% true positive rate. This is because the outliers will generally receive much lower depth values than the inliers, thus the only scenario in which the outliers are misclassified as inliers is when $k$ is large so that the whisker surpasses the outliers.

\begin{figure}[H]
    \centering
    \includegraphics[width=0.85\textwidth]{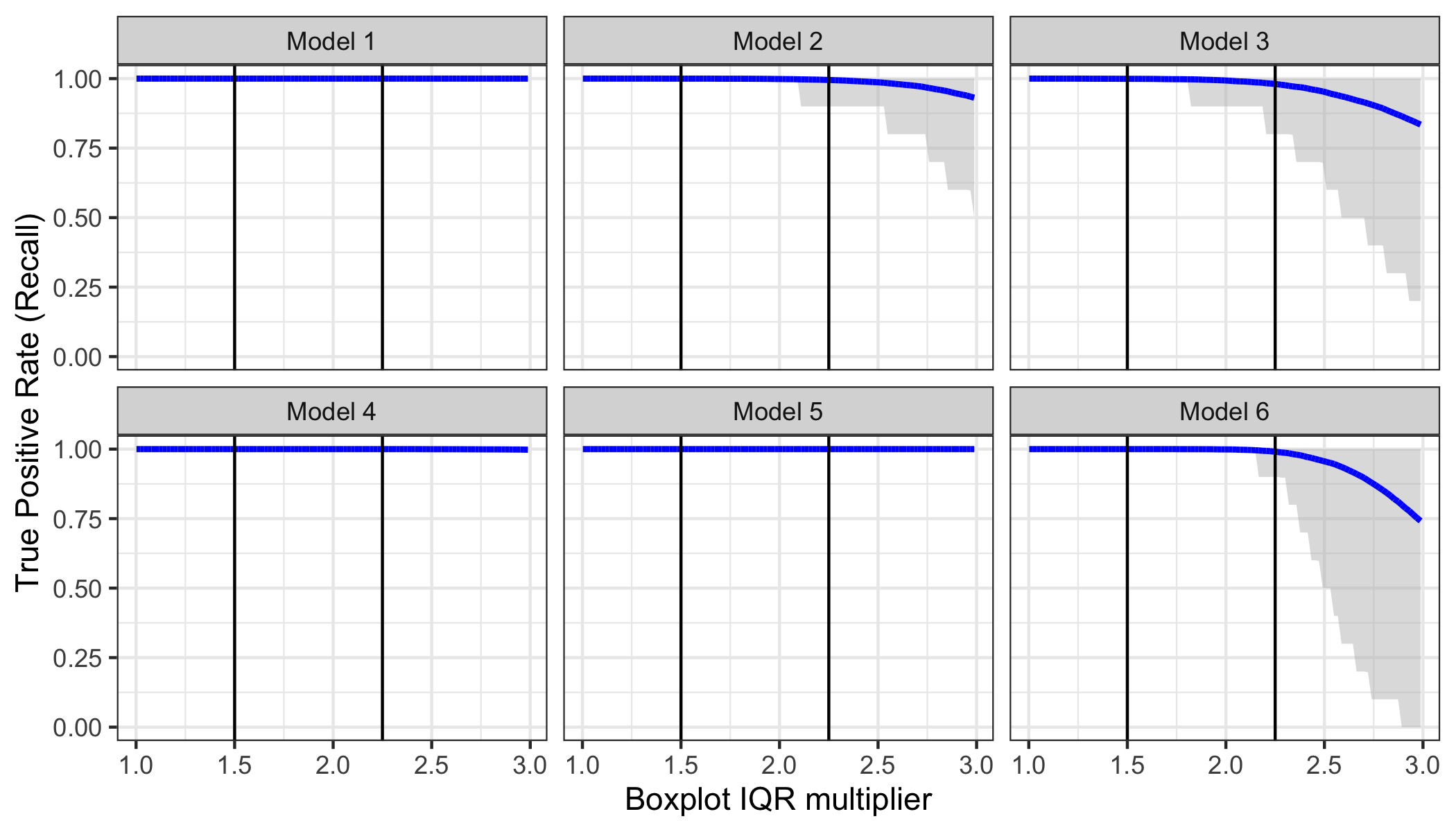}
    \caption{True Positive Rate (TPR) of the Amplitude depths on all amplitude models for boxplot multipliers $k = $ 1 to 3. Blue lines are the estimated mean and grey regions show the 95\% confidence interval. Vertical black bars indicate the upper and lower bounds of the suggested cutoff regions (1.5 to 2.25).}
    \label{fig:tpr}
\end{figure}

We also looked at how the coverage of our boxplots changed as a function of $k$ (Figure \ref{fig:tpr}). We ran the simulations again for the 6 amplitude models in the paper, except this time we allowed the boxplot multiplier $k$ to vary from 1 to 3. We measured the whisker's coverage using the True Negative Rate (TNR), which is the percentage of inliers classified as inliers.
    
Overall, we found the coverage (TNR) of the whisker to be fairly insensitive to the value of $k$. Across each model, the TNR steadily rises from about TNR = 0.95 (slight under coverage) at $k = 1$ until it asymptotes to TNR = 1 as $k$ approaches 3.    
    
\begin{figure}[H]
\centering
\includegraphics[width=0.85\textwidth]{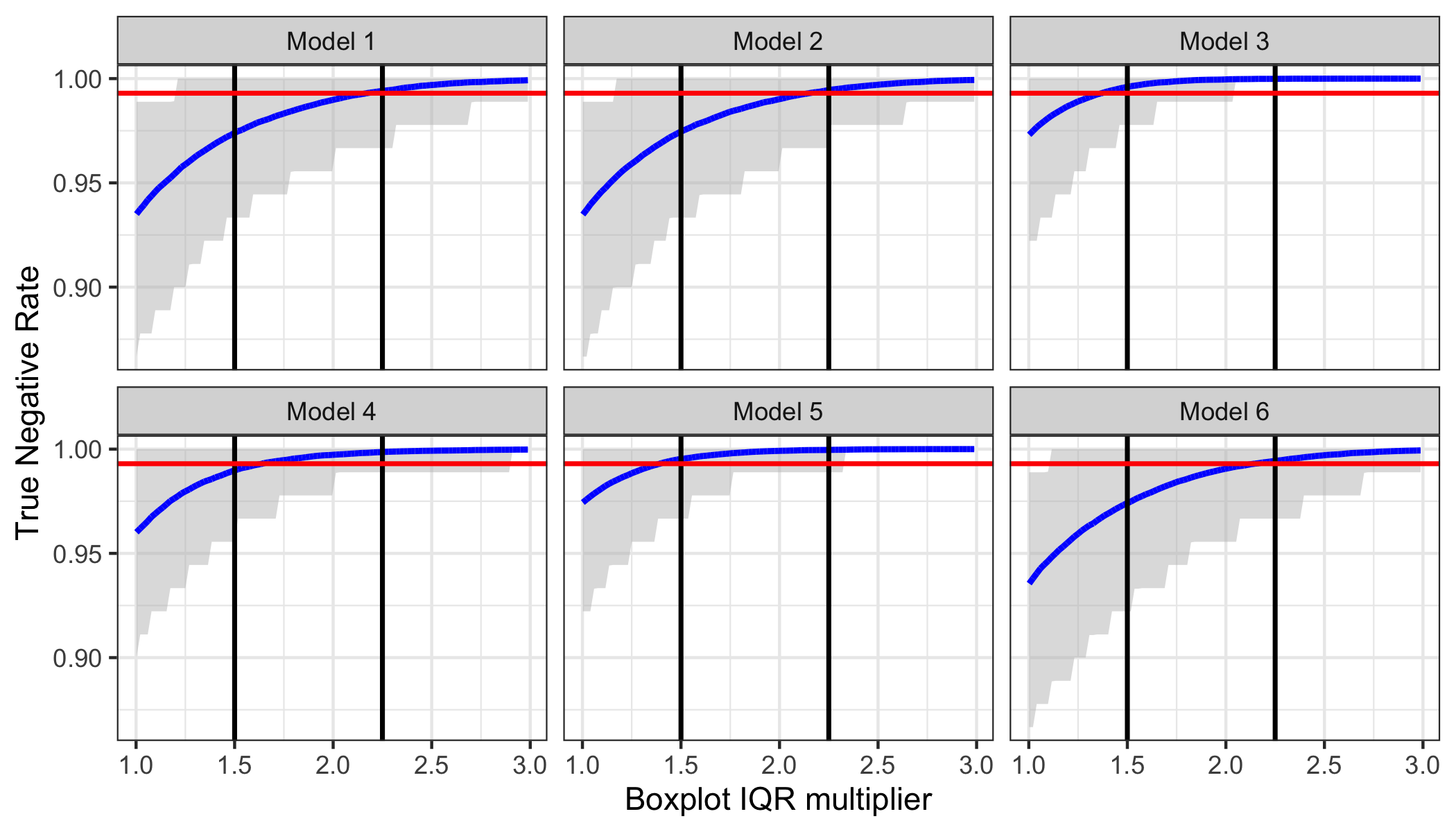}
\caption{True Negative Rate (TNR) of the Amplitude depths on all amplitude models for boxplot multipliers $k = $ 1 to 3. Blue lines are the estimated mean and grey regions show the 95\% confidence interval. Vertical black bars indicate the upper and lower bounds of the suggested cutoff regions (1.5 to 2). The red horizontal line indicates the desired TNR of 99.3\%. Proper coverage is achieved at the value of $k$ where the blue and red line cross.}
\label{fig:coverage}
\end{figure}
    
Models 1, 2, and 6 required larger values of $k$, than models 3, 4, and 5 to achieve the desired coverage of 99.3\%. However, we found this mostly had to do with the sampling rate of the observed trajectories. Essentially the trajectories in Models 1, 2, and 6 were undersampled relative to their frequency content (complexity) which caused the depth distribution to become left skewed and the boxplot to undercover the distribution.
    
\begin{figure}[H]
\centering
\subfloat{%
        \includegraphics[width=0.45\textwidth]{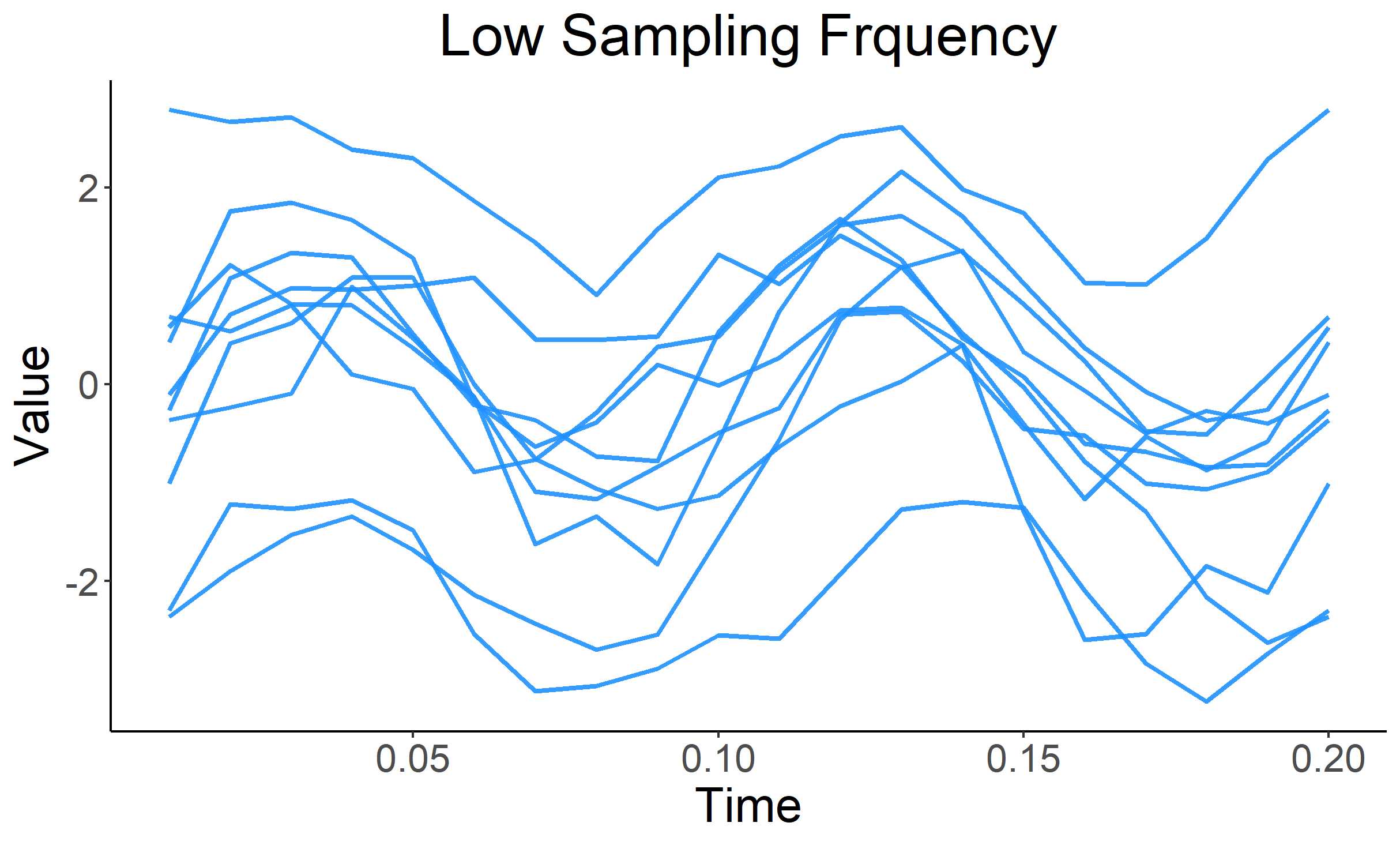}
        } 
        \subfloat{%
        \includegraphics[width=0.45\textwidth]{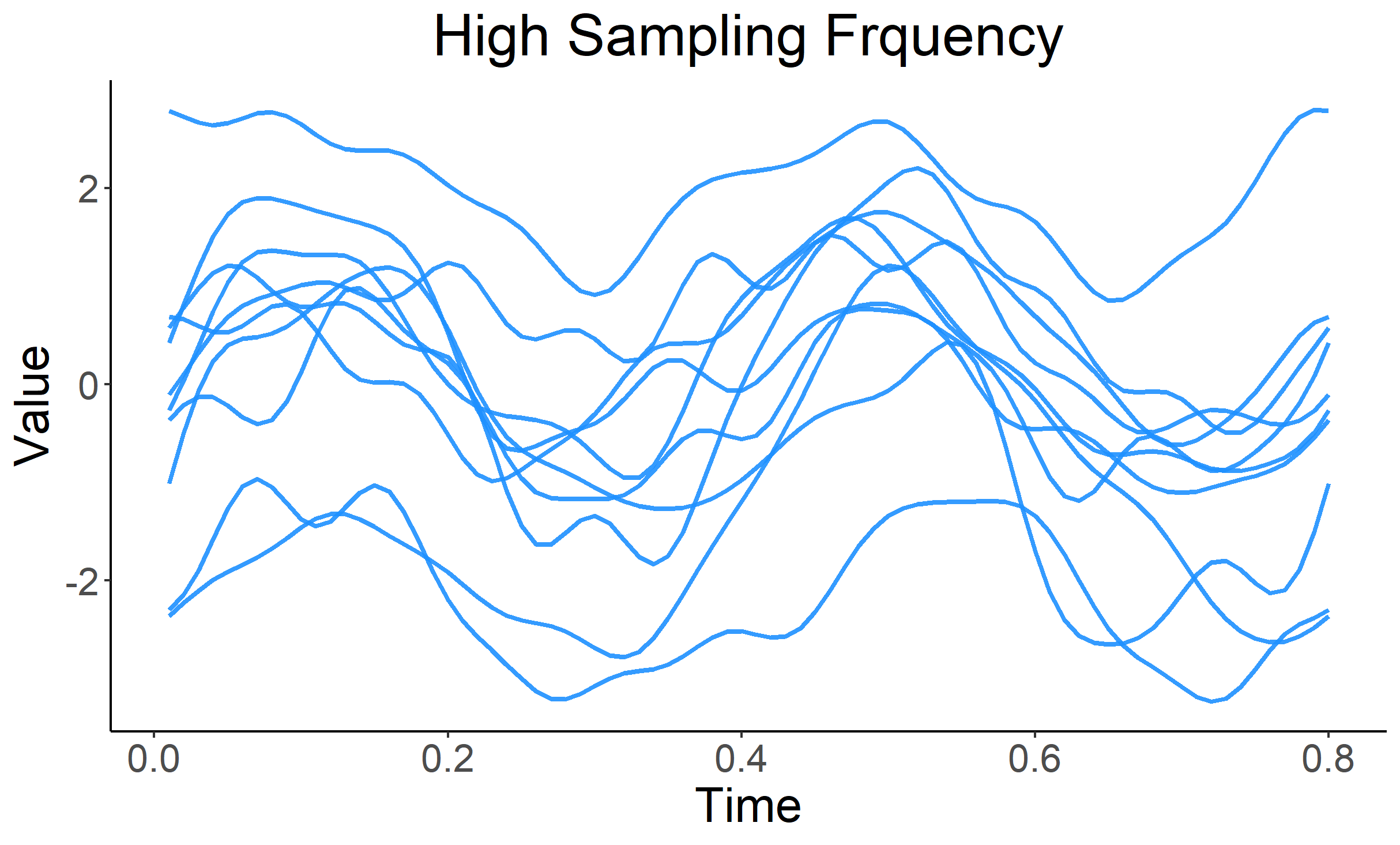}
        } \\
        \subfloat{%
        \hspace{3em} \includegraphics[width=0.7\textwidth]{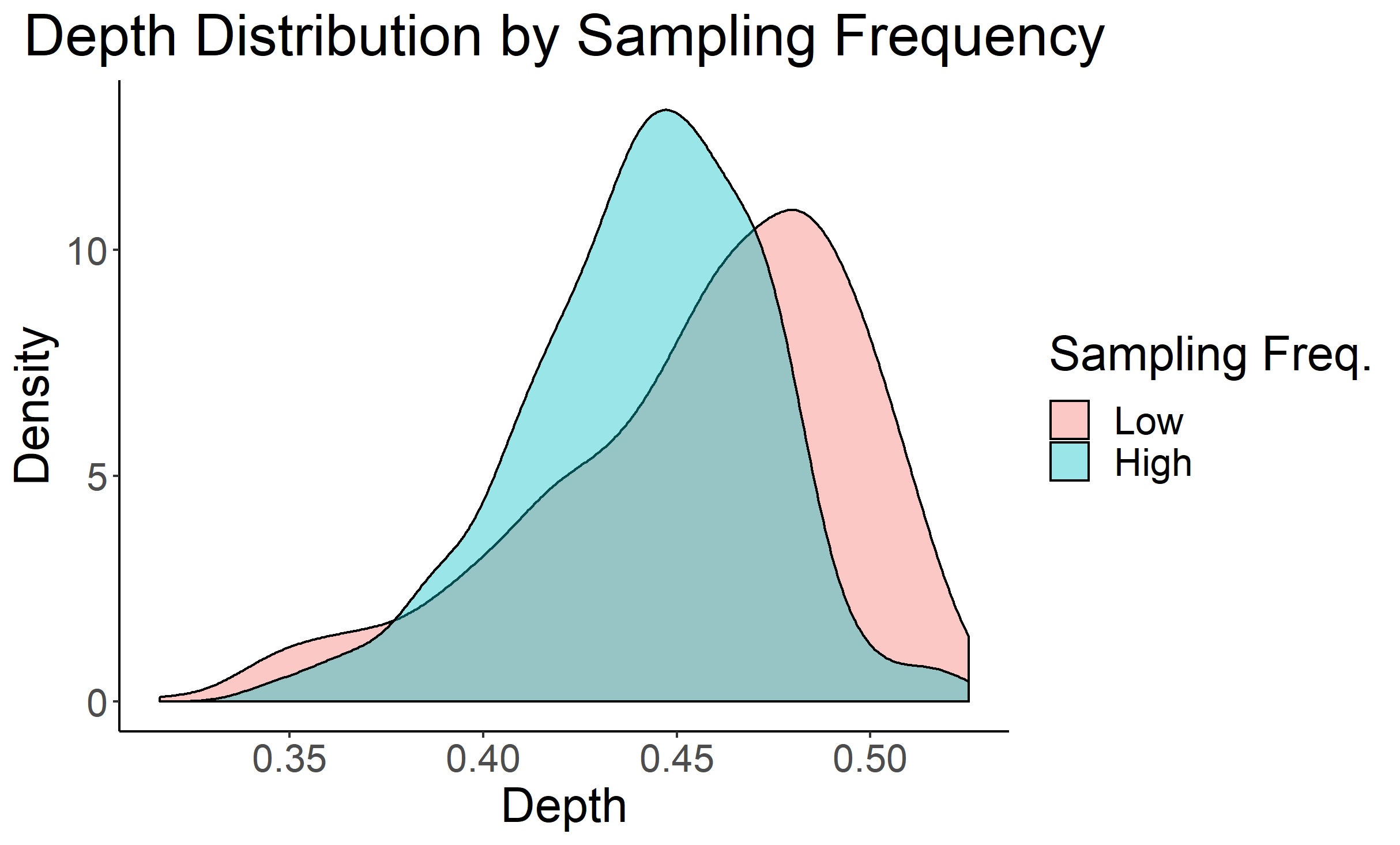}
        }
        \caption{\textbf{Top Left:} 10 functions with 21 frequency components, sampled at 20 points along $[0, 1]$ \textbf{Top Right:} The same functions sampled at 80 points along $[0, 1]$. \textbf{Bottom:} Distribution of the elastic depths at both sampling frequencies. Higher sampling frequency results in a more symmetric distribution which means a smaller $k$ is sufficient to achieve 99.3\% coverage.}
    \label{fig:sampling}
\end{figure}

When trajectories are sampled below their Nyquist rate, for example, twice the maximum frequency component present in the trajectory, they are considered under-sampled and aliasing occurs. This leads to poor alignment, which has the net effect of skewing the amplitude depth distribution due to underestimating many of the shape distances. In such a case, as long as the sampling rate is increased, then the amplitude depth distribution will become more symmetric and a lower value of $k$ can be used to achieve 99.3\% coverage. We demonstrate this in Figure \ref{fig:sampling}.
    
\subsection{Additional phase simulations} \label{sec:phase_appendix}
In Section 5.2 we noted that our phase simulations did not optimally represent pure phase differences. In fact, the phase outliers were more outlying in amplitude space than in phase space. To study pure phase differences better, we compared our method ED-P against TVD on a simple sin waves vs cosine waves example. 

\begin{figure}[H]
    \centering
    \includegraphics[width=0.85\textwidth]{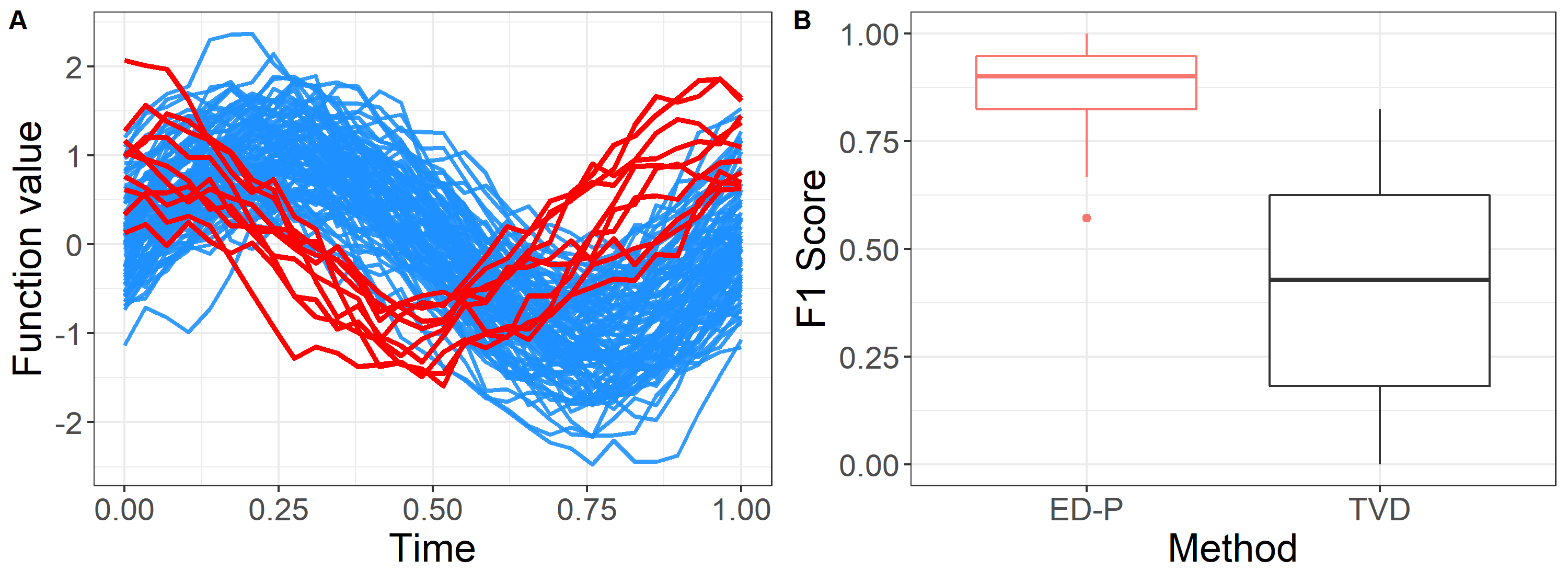}
    \caption{Panel A: sin waves (blue -- inliers) v.s. cosine waves (red -- outliers). Panel B: F1 score boxplots for ED-P and TVD when either method is used to detect the outliers.}
    \label{pa:fig:phase_comp}
\end{figure}

We generated inliers from a GP with a $\sin(2\pi t), t \in [0, 1]$ mean function and outliers from a GP with a $\cos(2\pi t)$ mean function. We used the exponential covariance function with $r = 0.5$ for the GPs covariance functions. 100 replications were performed to generate panel B of Figure \ref{pa:fig:phase_comp}. TVD's detection skill deteriorated significantly from the results in the manuscript, while ED-P did not. 
    
This result demonstrates the TVD was relying on non-phase differences to help differentiate phase outliers. Because, if TVD was not using non-phase information, then the loss of non-phase information would not have impacted its $F_1$ scores. 

\subsection{Examples} \label{sec:examples_appendix}
To demonstrate the real-world performance of elastic depth we applied it to three different data sets, one for each manifold. The first data set is a collection of U.S. Treasury yield curves ($\R$), the second is a sample from the MPEG-7 image data set ($\R^2$), and the last is hurricane trajectories across the Atlantic Ocean ($\mathbb{S}^2$). We show that the elastic depths can be applied consistently across each of the three manifolds and analyzed using the same boxplot methodology.

\subsection{U.S Daily Treasury Yield Curve Rates ($\mathbbm{R}$)} \label{bond}
We first consider the daily U.S. Treasury yield curves from January 2017 to April 2019 \citep{treasury}. The daily yield curve is a plot of bond terms, or time to maturity, against the associated interest rate on a given day.
The shape of the yield curve has long been taken as an indicator of economic activity. Ordinarily, the yield curve is monotonically increasing as a function of time to maturity. However, in the months preceding recessions, the yield curve often becomes ``flattened'' and then ``inverted'' meaning that bonds with shorter maturity dates start to command higher interest rates than bonds with longer maturity dates.

\begin{figure}
\centering
  \subfloat{%
    \includegraphics[width=0.45\textwidth,valign=c]{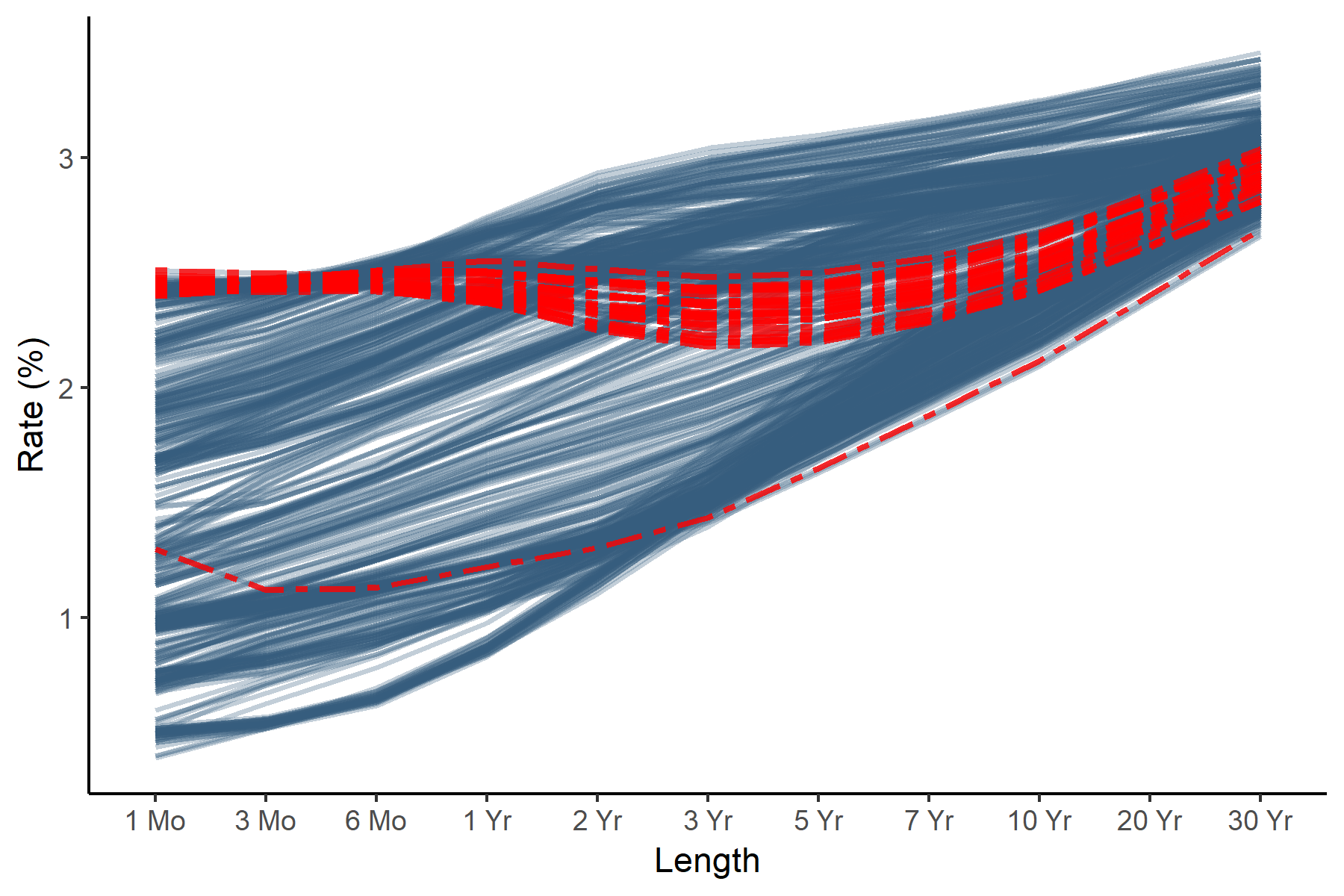}
  }
  \subfloat{%
    \includegraphics[width=0.45\textwidth,valign=c]{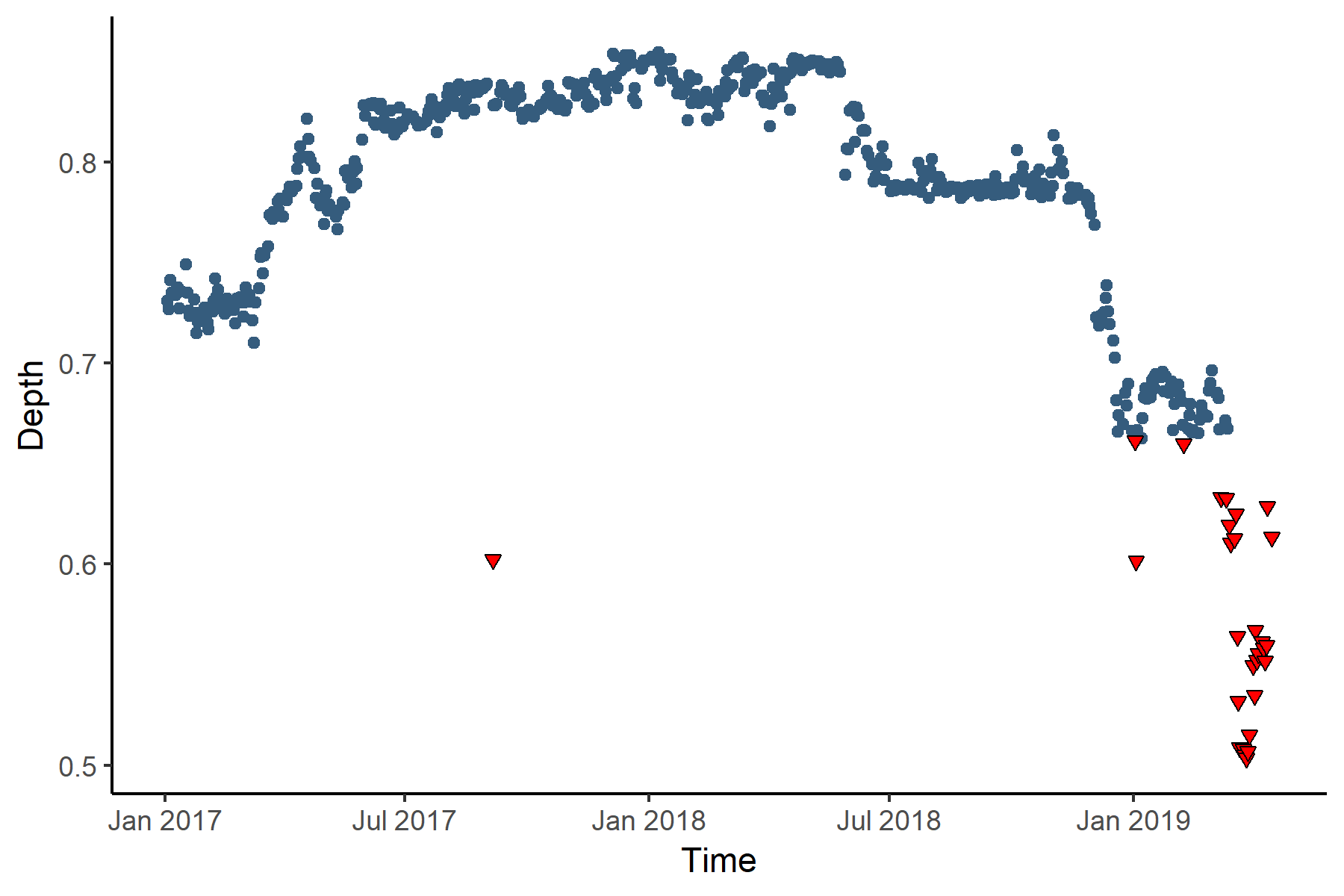}
  }
  \caption{\textbf{Left:} Shape (amplitude) outlying daily U.S. Treasury yield curves (dashed lines) against all daily yield curves from Jan. 2017 to Apr. 2019. \textbf{Right:} Amplitude depths for each curve plotted over time. The bulk of the outlying curves ( triangles), i.e. ``flat'' curves occur during the end of the observation period.}
  \label{yields}
\end{figure}

We collected yield curve data for each day between January 1st, 2017 to April 2019 with terms spanning from 1 month to 30 years.
Though the yield curve is only supported on a finite set of points, we treat it as though it were a continuous trajectory supported on a compact subset of $\R$. This assumption is not altogether unreasonable given the relatively smooth relationship between time to maturity and interest rate. We then applied the elastic depth based boxplots with no thresholding to perform outlier detection, see Figure \ref{yields}.

There are two sets of outliers identified in Figure \ref{yields}. The first is the yield curve on 09/05/17 where the 1 Month interest rate spiked over the 3 Month interest rate. Though the spike was enough to force that day's yield curve to become outlying, the overall shape of the curve is still monotonic, so this type of outlier is unlikely to be of practical interest. The second set of outliers is the group of 26 yield curves, also highlighted in red, that corresponds to the end of the observation period. Though these curves are not complete inversions they are considered ``flat'' since the 3 Month and 30 Year interest rates are similar. Because these flat curves follow a period of regular yield curve behavior they may be taken as a sign of an oncoming yield curve inversion and potentially an economic recession.

\subsection{MPEG-7 Shape Data ($\mathbbm{R}^2$)} \label{mpeg7}
Our next data example comes from the MPEG-7 shape data set \citep{manjunath2002introduction}. This data set consists of 1300 trajectories in $\R^2$, corresponding to 65 shape classes with 20 observations each. Each trajectory is an outline of some object such as an apple, turtle, or a butterfly sampled from frames in a video. Objects may be rotated, distorted, or magnified with respect to other objects of the same class.

\begin{figure}
\centering
  \subfloat{%
    \includegraphics[width=0.45\textwidth,valign=c]{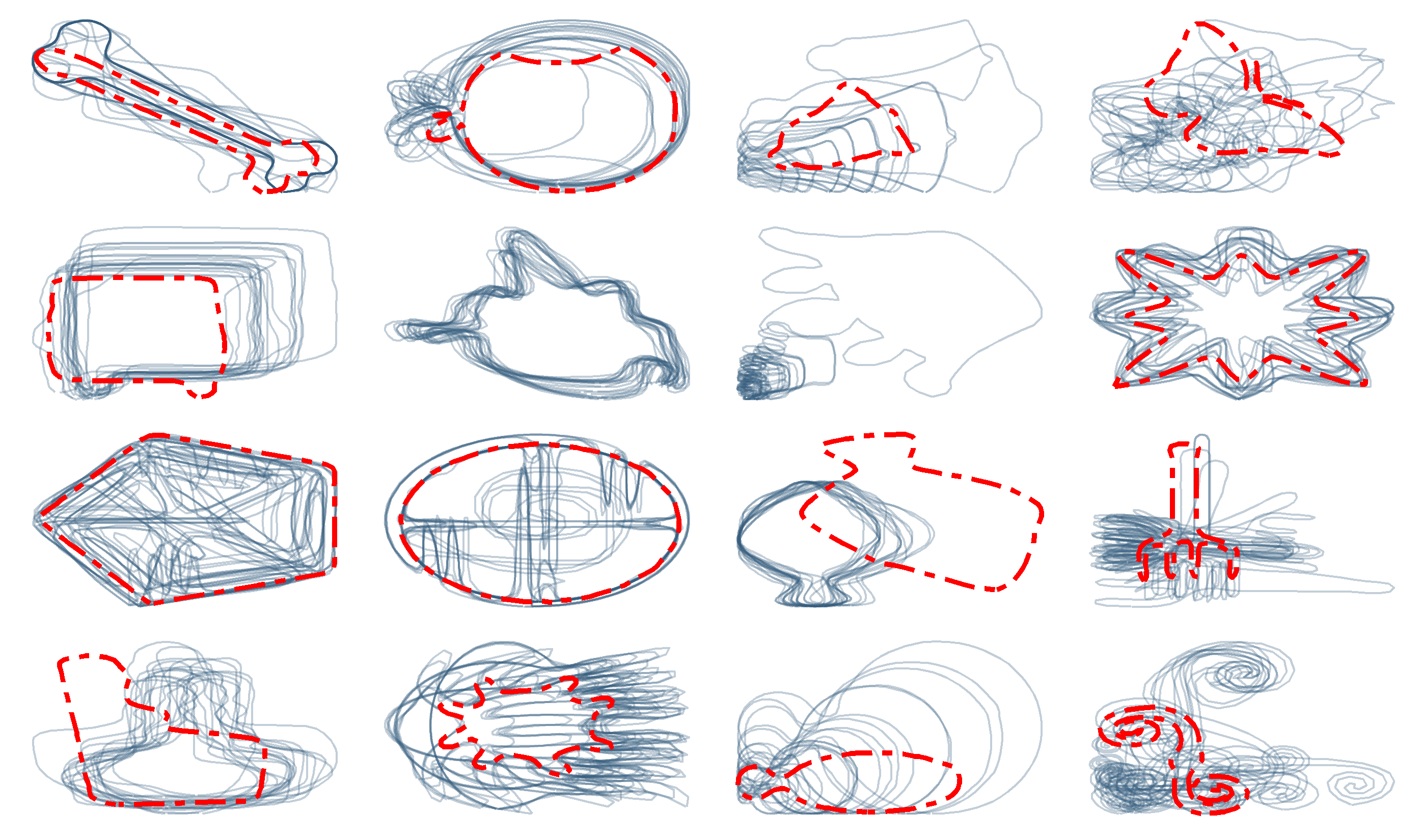}
  } \hspace{1cm}
  \subfloat{%
    \includegraphics[width=0.45\textwidth,valign=c]{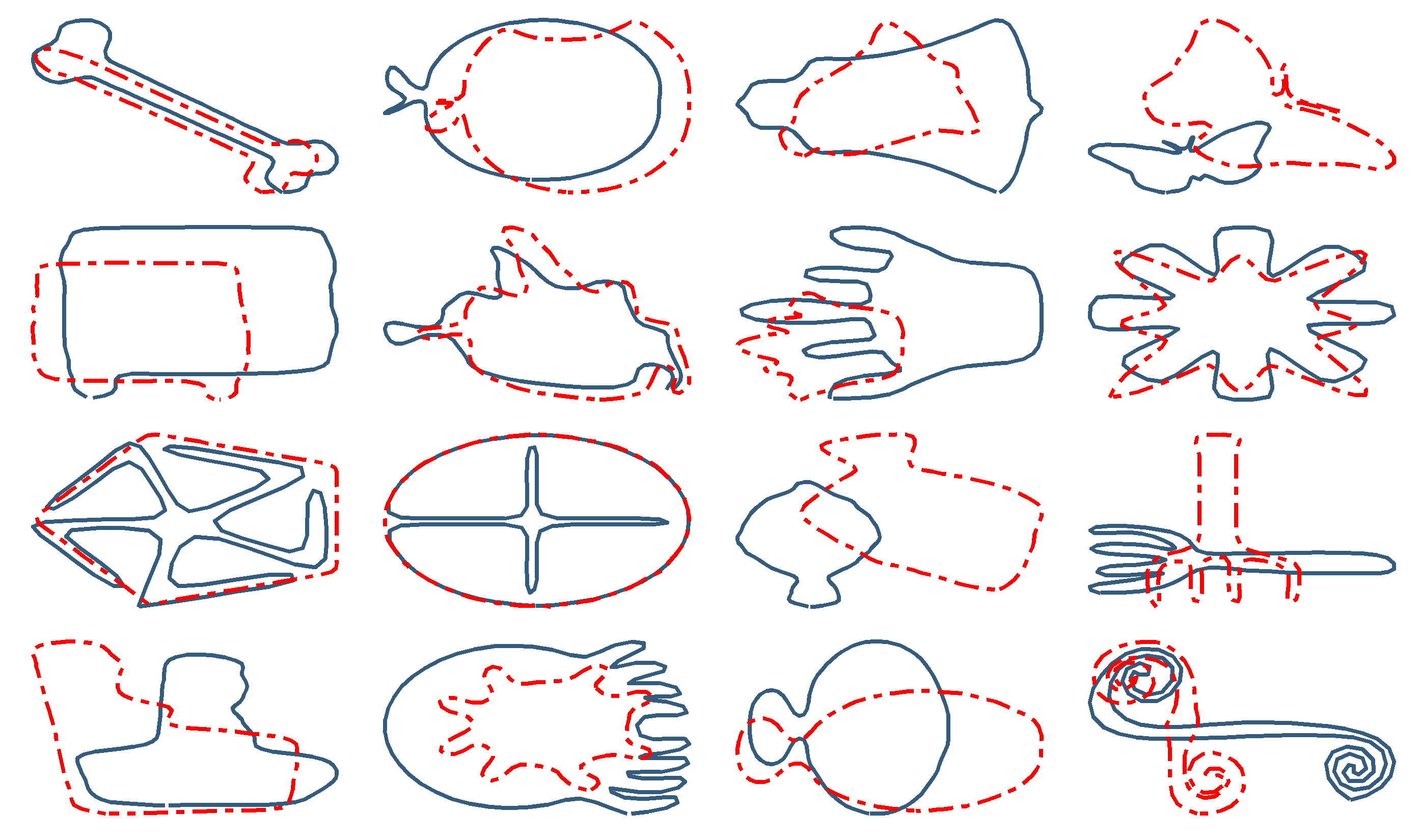}
  }
  \caption{\textbf{Left:} 16 shape classes from MPEG-7 along with their amplitude outliers (dashed lines). \textbf{Right:} Most central function in each shape class (solid line) overlaid with the most outlying function (dashed line) for each of the 16 classes.}
  \label{r2:outliers}
\end{figure}

To illustrate our method, we selected 16 shape classes from the available 65. We used a depth threshold of 0.05 so that at most one outlier would be detected within each shape class. The identified amplitude outliers are displayed in red against the inlying shapes in blue on the left hand side of Figure \ref{r2:outliers}.
We can see that, depending on the class, elastic depth can identify anomalous trajectories with both too high and too low amplitudes with respect to the rest of the set. For instance, the outlying jellyfish in the second column of row four has its tentacles spreading out from all sides of its body, whereas the inlying jellyfish all have their tentacles on the right. On the other hand, the star in the fourth column of row two lacks the larger rounded features of the inlying set which is why it too was identified as outlying.

In some of the classes, such as the bone or the pentagon, it is not visually obvious why the identified outliers are outlying in shape. 
To understand why these trajectories may be shaped differently than the rest, we plotted the most outlying trajectory (red) against the most central trajectory (blue) for each class in the right hand side of Figure \ref{r2:outliers}. This contrast makes clear that the bone outlier lacks a protrusion on the top and the pentagon outlier lacks spokes, features which would make them far in amplitude from other trajectories in their class. In other classes, such as the butterfly, there is no single feature that distinguishes the outliers so its outlyingness results from the culmination of numerous small shape differences.

\end{document}